\documentclass[fleqn,usenatbib,useAMS]{mnras}

\usepackage{graphicx}	
\usepackage{amsmath}	
\usepackage{amssymb}	
\usepackage{multicol}        
\usepackage{bm}		
\usepackage{pdflscape}	

\usepackage[T1]{fontenc}
\usepackage{ae,aecompl}

\usepackage{newtxtext,newtxmath}

\title[Radio Emission from White Dwarfs in VLASS]{A Survey for Radio Emission from White Dwarfs in the VLA Sky Survey}

\author[Pelisoli et al.]{
Ingrid Pelisoli$^{1}$\thanks{E-mail: ingrid.pelisoli@warwick.ac.uk}, Laura Chomiuk$^{2}$, Jay Strader$^{2}$, T.~R. Marsh$^{1}$,
Elias Aydi$^{2}$, Kristen C. Dage$^{3}$,\newauthor Rebecca Kyer$^{2}$, Isabella Molina$^{2}$, Teresa Panurach$^{2,4}$, Ryan Urquhart$^{2}$,
Thomas J. Maccarone$^{5}$, \newauthor R. Michael Rich$^{6}$, Antonio C. Rodriguez$^{7}$,
E. Breedt$^{8}$,
A.~J. Brown$^{9,10}$,
V.~S.\ Dhillon$^{10,11}$, \newauthor
M.~J. Dyer$^{10}$,
Boris. T. Gaensicke$^{1}$,
J.~A. Garbutt$^{10}$,
M.~J. Green$^{12,13}$,
M.~R. Kennedy$^{14}$,
P. Kerry$^{10}$,\newauthor
S.~P. Littlefair$^{10}$,
James Munday$^{1,15}$,
S.~G. Parsons$^{10}$
\\
$^{1}$Department of Physics, University of Warwick, Gibbet Hill Road, Coventry, CV4 7AL, UK\\
$^{2}$Center for Data Intensive and Time Domain Astronomy, Department of Physics and Astronomy, Michigan State University, East Lansing, MI 48824, USA\\
$^{3}$Wayne State University, Department of Physics \& Astronomy, 666 W Hancock St, Detroit, MI 48201, USA\\
$^{4}$Norfolk State University, 700 Park Avenue, Norfolk, VA 23504 USA\\
$^{5}$1Department of Physics \& Astronomy, Texas Tech University, Box 41051, Lubbock, TX, 79409-1051, USA\\
$^{6}$Department of Physics and Astronomy, University of California, Los Angeles, CA, 90095, USA\\
$^{7}$Department of Astronomy, California Institute of Technology, 1200 E. California Blvd, Pasadena, CA, 91125, USA\\
$^{8}$Institute of Astronomy, University of Cambridge, Madingley Road, Cambridge CB3 0HA, UK\\
$^{9}$Departament de F\'{\i}sica, Universitat Polit\`{e}cnica de Catalunya, c/Esteve Terrades 5, 08860 Castelldefels, Spain\\
$^{10}$Department of Physics and Astronomy, Hicks Building, The University of Sheffield, Sheffield, S3 7RH, UK\\
$^{11}$Instituto de Astrof\'\i{}sica de Canarias (IAC), E-38205 La Laguna,  Tenerife, Spain \\
$^{12}$Max-Planck-Institut f\"{u}r Astronomie, K\"{o}nigstuhl 17, D-69117 Heidelberg, Germany\\
$^{13}$Department of Astrophysics, School of Physics and Astronomy, Tel Aviv University, Tel Aviv 6997801, Israel\\
$^{14}$Department of Physics, University College Cork, Cork, Ireland\\
$^{15}$Isaac Newton Group of Telescopes, Apartado de Correos 321, Santa Cruz de La Palma, E-38700, Spain\\
}
\date{Last updated XXX; in original form XX}

\pubyear{2024}

\begin{document}
\label{firstpage}
\pagerange{\pageref{firstpage}--\pageref{lastpage}}
\maketitle

\begin{abstract}
Radio emission has been detected from tens of white dwarfs, in particular in accreting systems. Additionally, radio emission has been predicted as a possible outcome of a planetary system around a white dwarf. We searched for 3 GHz radio continuum emission in 846,000 candidate white dwarfs previously identified in {\it Gaia} using the Very Large Array Sky Survey (VLASS) Epoch 1 Quick Look Catalogue. We identified 13 candidate white dwarfs with a counterpart in VLASS within 2". Five of those were found not to be white dwarfs in follow-up or archival spectroscopy, whereas seven others were found to be chance alignments with a background source in higher-resolution optical or radio images. The remaining source, WDJ204259.71+152108.06, is found to be a white dwarf and M-dwarf binary with an orbital period of 4.1~days and long-term stochastic optical variability, as well as luminous radio and X-ray emission. For this binary, we find no direct evidence of a background contaminant, and a chance alignment probability of only $\approx 2$ per cent. However, other evidence points to the possibility of an unfortunate chance alignment with a background radio and X-ray emitting quasar, including an unusually poor Gaia DR3 astrometric solution for this source. With at most one possible radio emitting white dwarf found, we conclude that strong ($\gtrsim 1-3$~mJy) radio emission from white dwarfs in the 3~GHz band is virtually nonexistent outside of interacting binaries.
\end{abstract}




\begingroup
\let\clearpage\relax
\endgroup
\newpage

\section{Introduction}
\label{intro}

White dwarf stars are the most common remnant of stellar evolution, with over two billion of them predicted to exist in our Galaxy \citep{Nelemans+2001}. Thanks to the precise data from the {\it Gaia} data release 3 \citep{GaiaDR3}, around 1.3~million white dwarf candidates have been identified \citep{GentileFusillo+21}.
The abundance of white dwarfs compared to other stellar fossils stems from the fact that all stars with masses below 7--10.6~M$_{\odot}$ become white dwarfs \citep{WoosleyHeger2015,Lauffer+2018} and, due to the steepness of the initial-mass function \citep[e.g.][]{Kroupa2001,Padoan+2002}, that corresponds to the vast majority of stars ($>90$\%). This makes them excellent laboratories for studying stellar evolution.

Around a quarter of white dwarfs are found in binary systems, half of which with main sequence companions \citep{Toonen+2017}. For orbital periods shorter than at least 6~hours, cool main sequence companions will fill their Roche lobe and transfer mass onto the white dwarf. These accreting white dwarfs are called cataclysmic variables \citep[CVs, see][for a thorough review]{Warner1995}. Cataclysmic variables are classified into different types depending on the observed properties of their optical light curves, as well as on the strength of the white dwarf's magnetic field. For weak or absent magnetic fields ($\lesssim 1$~MG), the mass transferred from the companion will form an accretion disk. Instabilities in this disk can lead to changes in the mass transfer rate, leading to strong outbursts during which the observed optical magnitude can increase by two orders of magnitude on timescales of weeks to decades. Systems that exhibit this are called dwarf-novae. 
Systems with high mass transfer rates where outbursts are not observed are called nova-likes \citep[e.g.][]{Dubus+2018}. For some CVs, brightness changes of over six magnitudes are observed, which are attributed to a thermonuclear explosion on the surface of the white dwarf; these are called classical novae. When the white dwarf's magnetic field is strong, the accretion disk will be either fully suppressed (polars, $B \gtrsim 10$~MG), or truncated at the white dwarf magnetosphere (intermediate polars, or IPs, $1 \lesssim B \lesssim 10$~MG), and accretion will follow the magnetic field lines and occur at the magnetic pole.

Radio emission is commonly observed in CVs, in particular for magnetic systems, but has also been reported for non-magnetic systems. There are 45 CVs (33 of which are magnetic) reported as radio detections in the literature, though some have low significance \citep[see Table~\ref{tab:radiodets} and][]{Kording2008, Coppejans2015, Coppejans2016, Russell2016, Barrett2020, Hewitt2020, Ridder2023}. In contrast, radio emission in the absence of accretion seems to be a rare feature of white dwarfs and has only been detected in two types of systems: magnetic propellers and binary white dwarf pulsars. In both cases, the white dwarf has a relatively close binary companion but no transferred material reaches the white dwarf surface. Only two systems of each kind are known. In the magnetic propellers---AE~Aquarii \citep[AE~Aqr;][]{Patterson1979,Bookbinder1987} and LAMOST~J024048.51+195226.9 \citep[J0240+1952;][]{Thorstensen2020,Pretorius+2021,Pelisoli+2022a}---mass is being transferred towards the white dwarf from the companion, but the material is then flung out by the white dwarf's magnetic field in synchrotron-emitting blobs \citep{Bastian1988,Meintjes2005}. In the binary white dwarf pulsars---AR Scorpii \citep{Marsh2016} and J191213.72$-$441045.1 \citep[J1912$-$4410;][]{Pelisoli+23, Schwope2023}---both the spectral features and the observed light curves suggest an absence of steady accretion. The proposed explanation is that these systems are experiencing a detached phase due to transfer of angular momentum from the white dwarf spin onto the orbit \citep{Schreiber2021}. For binary white dwarf pulsars, we observe pulsed emission, from radio to X-rays, that varies over the spin period of the white dwarf as the white dwarf receives an injection of electrons when its magnetic field sweeps past the companion. Different locations for the exact origin of the emission have been proposed, including the magnetosphere \citep{Takata2017, PotterBuckley2018}, the white dwarf surface \citep{Plessis2022}, and the surface or coronal loops of the M-dwarf companion \citep{Katz2017}. For J1912$-$4410, the radio pulses are remarkably narrow, in contrast with the broad pulses observed for AR~Sco, and resemble neutron star pulsar features, suggesting that beamed synchrotron emission is perhaps being observed directly from the white dwarf magnetic pole.

The observed behaviour of these radio-emitting white dwarfs, and in particular of J1912$-$4410, raises the question of whether rapidly spinning magnetic white dwarfs could produce pulsar-like radio emission, driven
by the spin-down of the magnetic dipole \citep[e.g.][]{Harayama+2013}. Another possibility for producing radio
emission from white dwarfs is with orbiting planets: planets may induce a current, driven by the white dwarf magnetic field and producing electron-cyclotron maser emission, in analogy with the Jupiter–Io
interaction \citep{Li1998, WillesWu2004, WillesWu2005}. This list of possible mechanisms for producing radio emission from white dwarf is not necessarily comprehensive, and there could be others not yet proposed or considered.

With these possibilities in mind, in this work we searched for radio counterparts to the white dwarf candidates catalogued by \citet{GentileFusillo+21} in the Very Large Array Sky Survey (VLASS) Epoch 1 Quick Look Catalogue \citep{Gordon+20, Gordon+21}. We inspected images of each matched source to exclude chance alignments with background sources. Follow-up optical observations for sources where the radio emission matched the position of the white dwarf 
were carried out primarily with the Goodman Spectrograph \citep{goodman} on the 4.1~m Southern Astrophysical Research (SOAR) telescope, and with the fast photometers ULTRACAM \citep{ultracam}, mounted at the 3.5-m ESO New Technology Telescope (NTT), and ULTRASPEC \citep{ultraspec} at the 2.4-m Thai National Telescope (TNT).

\section{Selection of candidate radio-emitting white dwarfs}\label{sec:gf}

We searched for radio counterparts to 846,000 white dwarf candidates catalogued by \citet{GentileFusillo+21} that are in the VLASS footprint. {\citet{GentileFusillo+21} selected} candidates using \emph{Gaia} eDR3 data only, by doing colour-magnitude and quality control cuts defined based on known white dwarfs. Their selection is tailored for single white dwarf stars, including also white dwarfs with low-luminosity companions or double white dwarf binaries, which occupy the same colour-magnitude region as single white dwarf stars, but does not aim at including white dwarfs with main sequence companions (such as CVs). Systems with poor astrometric solutions due to being in binaries whose on-sky separation is comparable to the {\it Gaia} resolution ($\approx 0.4''$), or due to photoemtric variability, might also be excluded. Yet, it is also the largest, most homogeneous, and widely used catalogue of white dwarf candidates, making it ideal to search for radio counterparts to white dwarfs. Overall, they estimated a completeness between 67 and 93 per cent for $G \leq 20$, $T_\mathrm{eff} > 7000$~K, and galactic longitude $|b| > 20^{\circ}$. Within 40~pc, the catalogue is 97 per cent complete \citep{OBrien2024}. Each candidate was assigned a white dwarf probability $P_\mathrm{WD}$ based on the location of candidates in the {\it Gaia} colour-magnitude diagram, with higher values assigned to objects closer to the single white dwarf cooling sequence. We note that, as a consequence, binary white dwarf systems, which are common radio emitters, can show low $P_\mathrm{WD}$, and hence we make no selection on $P_\mathrm{WD}$ in this work.

We cross-compared this \emph{Gaia} white dwarf catalogue with the VLASS Epoch 1 Quick Look Catalogue \citep{Gordon+20, Gordon+21}. VLASS is conducted using the Karl G. Janksy Very Large Array telescope \citep[VLA,][]{Perley2011} in the S-band (2--4~GHz). The survey covers all the sky above declination $-40^{\circ}$. Data acquisition started in September 2017, and is predicted to finish in 2024. The data are split into three epochs to allow for variability searches. Data quality is not homogeneous throughout the epochs, as the pipeline continues to be improved, but the typical noise level in quick look images is 120~$\mu$Jy \citep{Gordon+21}. Three known CVs were previously reported as detections \citep{Ridder2023} in VLASS images. The source catalogue of \citet{Gordon+20, Gordon+21} that we employ for our crossmatch was constructed using an automated source finder in the quick look images \citep[PyBDSF,][]{PyBDSF}, which resulted in 625,000 unique sources. The catalogue is complete down to 3~GHz flux densities $\geq$3~mJy. Out of the known white dwarf radio emitters, only AE~Aqr is identified in this source catalogue.

\begin{figure}
    \includegraphics[width=\columnwidth]{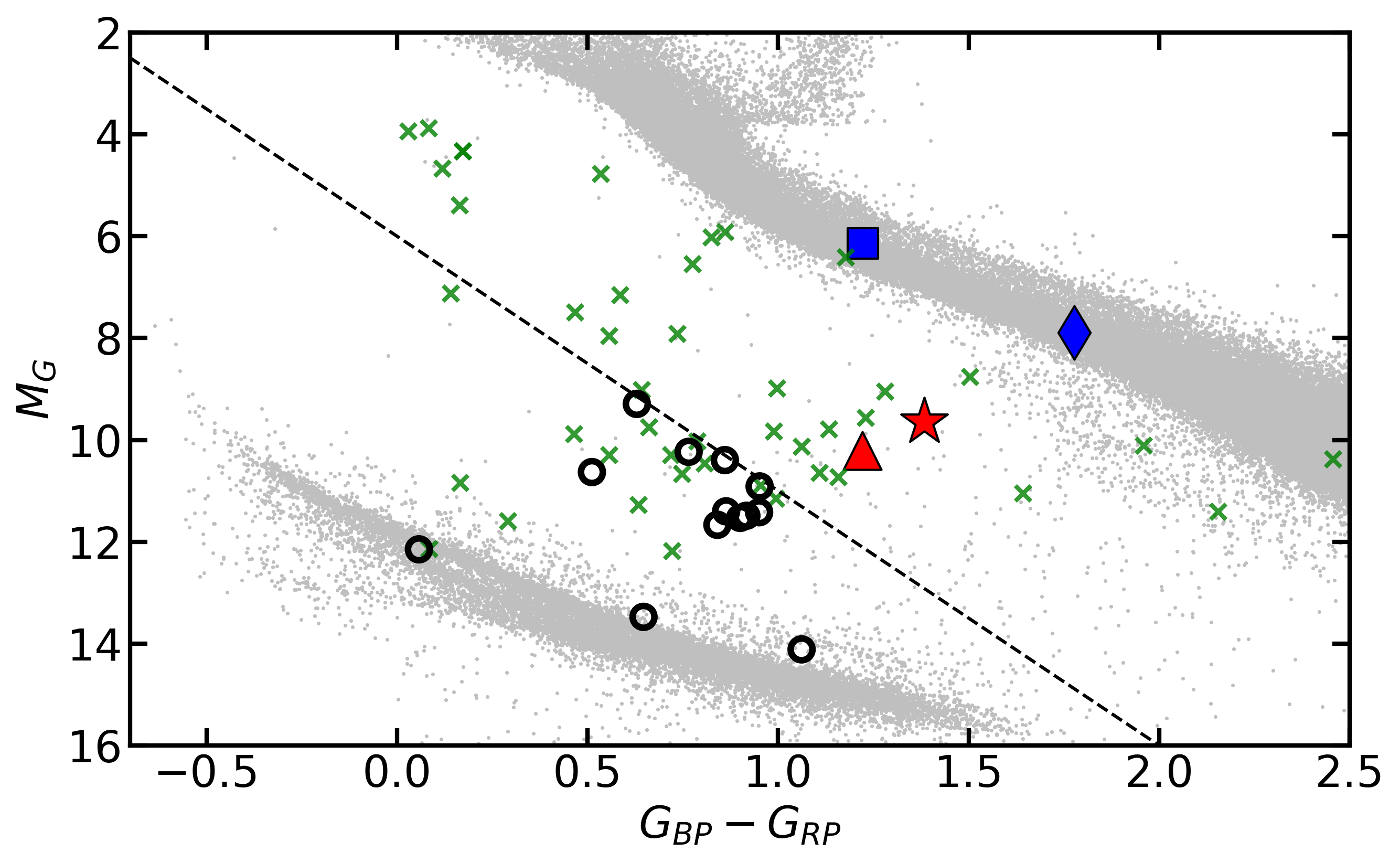}
    \caption{Absolute {\it Gaia} $G$ magnitude as a function of $G_{BP} - G_{RP}$ colour. The 13 white dwarf candidates with a 2$^{\prime\prime}$ match in VLASS are shown as open black circles. \citet{GentileFusillo+21} selected objects below the black dashed line. It excludes the location of the magnetic propellers shown as blue symbols (AE~Aqr: square, J0240: diamond) and also the location of the binary white dwarf pulsars shown in red (AR~Sco: star, J1912: triangle) due to their proximity to the main sequence. The CVs known to be radio emitters listed in Table~\ref{tab:radiodets} are shown as green crosses. The grey dots in the background are stars within 100~pc added to illustrate the location of the main sequence and white dwarf cooling track.}
    \label{fig:cmd}
\end{figure}

We used J2000 coordinates and initially used a generous search radius of 2$^{\prime\prime}$, which is significantly larger than the typical astrometric uncertainty of VLASS sources ($\sim0.5-1^{\prime\prime}$; VLASS Project Memo \#13\footnote{https://library.nrao.edu/public/memos/vla/vlass/VLASS\_013.pdf}). We chose this large radius to strive for completeness and reduce the chance of missing sources whose coordinates are uncertainty outliers. For similar reasons, we make no quality control cuts such as excluding extended sources at this stage. This resulted in 13 matches, listed in Table~\ref{tab:wd_vlass}, whose locations in the {\it Gaia} colour-magnitude diagram are shown in Figure~\ref{fig:cmd}. This figure also illustrates that the selection of \citet{GentileFusillo+21} excludes some known radio emitters, such as the magnetic propellers and binary white dwarf pulsars. It does, however, extend into the colour range occupied by some CV subclasses, and 12 of the previously known radio-emitting CVs are in the catalogue.

To investigate the radio variability of these white dwarf--VLASS matches, we downloaded Quick Look images for VLASS Epoch 1 (observed during 2017--2019) and Epoch 2 (2020--2022) from the CIRADA Image Cutout Web Service\footnote{http://cutouts.cirada.ca/}.
Flux densities were measured for the VLASS images by fitting a Gaussian (fixed to the width of the synthesized beam, as expected for a point source) to the radio sources. Some sources looked like they could be extended in the VLASS images, but in all cases the extension was mild/marginal and therefore it was not used as a criterion for exclusion as a possible white dwarf source. It is worth noting that flux density values in the Quick Look images have been found to be systematically low by on average $\approx 10$ per cent \citep{Gordon+21}. Additionally, substantial scatter ($\approx 15$ per cent) was found in the ratio of flux densities measured by VLASS and pointed observations of calibrator targets \citep{Lacy+19}. Reported fluxes should therefore be interpreted with caution. The uncertainties that we quote here are statistical only and reflect the noise levels of the VLASS images, but these systematic uncertainties borne of issues with calibration will also affect the flux densities.

\begin{table*}
	\centering
	\caption{The 13 white dwarf candidates with a match in VLASS. The name, white dwarf probability $P_\mathrm{WD}$ and {\it Gaia} $G$ magnitude are as reported in \citet{GentileFusillo+21}. Names are in bold when no background source better matched with the radio coordinates was identified (also noted in the ``Background contamination" column). The "WD" prefix from \citet{GentileFusillo+21} is dropped for systems found not to be white dwarfs. Also listed are the 3 GHz flux densities measured in VLASS Epoch 1 (observed during 2017--2019) and Epoch 2 (2020--2022). The last column indicates the separation in arcseconds between the white dwarf candidate and the radio source.}
	\label{tab:wd_vlass}
	\begin{tabular}{ccccccc} 
		\hline
		Name & $P_\mathrm{WD}$ & $G$   & \multicolumn{2}{c}{Flux Density in VLASS [mJy]} & Background & Separation \\
  		     &         & [mag] &           Epoch 1 & Epoch 2               & contamination & [$^{\prime\prime}$] \\
		\hline
            WDJ032623.40-243623.72 & 0.21 & 20.686$\pm$0.007 & 1.38$\pm$0.12 & 1.75$\pm$0.13 & yes & 1.65\\
            \textbf{J083802.17+145802.98} & 0.24 & 20.651$\pm$0.010 & 3.15$\pm$0.12 & 2.88$\pm$0.16 & no & 0.50\\ 
            \textbf{J105211.93-335559.93} & 0.11 & 19.664$\pm$0.006 & 12.02$\pm$0.16 & 12.64$\pm$0.12 & no & 0.87\\ 
            \textbf{J105223.24+313012.81} & 0.84 & 20.564$\pm$0.012 & 1.22$\pm$0.12$^a$ & 0.84$\pm$0.12$^a$ & no & 0.40\\ 
            \textbf{J120358.90-001241.07} & 0.19 & 20.485$\pm$0.008 & 1.09$\pm$0.14 & 1.03$\pm$0.15 & no & 1.93\\ 
            \textbf{WDJ121604.90-281909.67} & 1.00 & 18.784$\pm$ 0.003 & 2.23$\pm$0.12 & 2.29$\pm$0.14 & no & 0.84\\ 
            \textbf{J124520.10-351755.53} & 0.24 & 20.656$\pm$0.008 & 3.70$\pm$0.13$^a$ & 3.32$\pm$0.14$^a$ & no & 0.19\\ 
            WDJ132423.32-255746.37 & 0.044 & 20.183$\pm$0.006 & 1.97$\pm$0.12 & 1.84$\pm$0.14 & yes & 1.85\\
            \textbf{WDJ182050.14+110832.09} & 0.97 & 20.044$\pm$0.007 & 12.04$\pm$ 0.15 & 16.84$\pm$0.13 & no & 0.89\\ 
            WDJ182112.17+204801.17 & 0.06 & 20.419$\pm$0.005 & 37.27$\pm$0.15$^a$ & 27.63$\pm$0.13$^a$ & yes & 1.71\\
            WDJ183758.54-330258.93 & 0.07 & 19.837$\pm$0.007 & 5.21 $\pm$ 0.12 & 4.71 $\pm$ 0.14 & yes & 1.58\\
            WDJ185250.55-310839.29 & 0.038 & 19.766$\pm$0.006 & 5.50 $\pm$ 0.14 & 5.18 $\pm$ 0.15 & yes & 1.35\\
            \textbf{WDJ204259.71+152108.06} & 0.12 & 16.877$\pm$0.004 & 10.36 $\pm$ 0.12 & 11.90 $\pm$ 0.18 & no & 0.40\\ 
  		\hline

        \end{tabular}
 
   {\raggedright $^a$This radio source appears mildly extended in the VLASS images, so this flux density (which assumes a point source) is a lower limit on the flux. \par}
\end{table*}

\subsection{Candidate vetting}

To further assess the validity of the white dwarf--VLASS matches, we compared the VLASS Quick Look images with deep optical images obtained by Dark Energy Camera (DECam) and Panoramic Survey Telescope and Rapid Response System \citep[Pan-STARRS,][]{ps1}. The DECam images, primarily processed as part of the DECam Local Volume Exploration survey \citep[DELVE,][]{delve}, were downloaded from the NOIRLab Astro Data Lab\footnote{\url{https://datalab.noirlab.edu/sia.php}}. The images' world coordinate systems were matched to \emph{Gaia} DR2 with 22 mas accuracy \citep{Drlica-Wagner+21}. The Pan-STARRS images were placed on the \emph{Gaia} DR1 reference frame with astrometric uncertainty of 6 mas \citep{Magnier+20}.

The comparison of the VLASS images with these deep optical images revealed that, for five matches, the radio position was more consistent with a nearby interloper (likely background galaxies). For all of those, the separation between the {\it Gaia} and VLASS coordinates was more than $1^{\prime\prime}$ (see the non-bold-faced rows in Table~\ref{tab:wd_vlass}). Only one of the sources with separation of more than $1^{\prime\prime}$ showed no obvious background contamination. We further consider this source, and the others with $<1^{\prime\prime}$ separations, in the next section.

\section{Follow-up characterisation}

\subsection{Spectroscopy}

The candidate vetting left eight candidate white dwarfs with matching radio sources. Considering that most of those were assigned low $P_\mathrm{WD}$ in \citet{GentileFusillo+21} (Table ~\ref{tab:wd_vlass}), the next step in our analysis was to obtain spectra to confirm the white dwarf nature of these candidates.

\begin{figure}
    \includegraphics[width=\columnwidth]{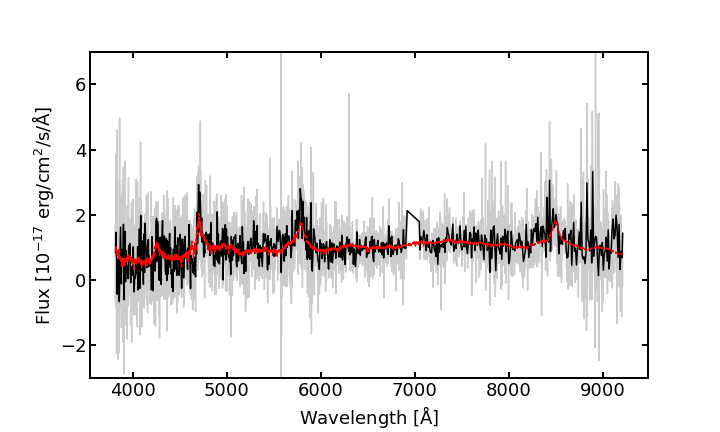}
    \caption{The SDSS spectrum of WDJ105223.24+313012.81 is shown in grey. The black line is a five-point average, and the red dashed line shows the best-fit SDSS template. The broad emission features are typical of quasars, and indicate a redshift of $z = 2.0346\pm0.0015$.}
    \label{fig:J1052spec}
\end{figure}

We started by querying the Sloan Digital Sky Survey (SDSS) data release 16 \citep[DR16][]{Ahumada2020} and the Large Sky Area Multi-Object Fibre Spectroscopic Telescope (LAMOST) data release 7 \citep[DR7;][]{Luo2022}. Only one match was found, WD~J105223.24+313012.81, which has an SDSS spectrum. The features in the spectrum (Fig.~\ref{fig:J1052spec}) clearly indicate that, despite the high white dwarf probability (Table~\ref{tab:wd_vlass}), WD~J105223.24+313012.81 is a quasar. The significant {\it Gaia} parallax of $5.1\pm1.2$~mas is likely the main cause for the high $P_\mathrm{WD}$, but it is worth noting that \citet{GentileFusillo+21} do classify this object as a quasar in their {\it Gaia}-SDSS spectroscopic sample. This object also appears as a quasar candidate in {\it Gaia} DR3. The fact that quasars show median parallaxes offset from the expected value of zero is well-known in {\it Gaia}; the mean offset is only of the order of $\mu$mas, but the standard deviation of the distribution of quasar parallaxes is over 1~mas \citep{Lindegren2021}.
WD~J105223.24+313012.81 seems to be one of the more extreme examples ($\approx 5\sigma$ from the mean) of a quasar parallax outlier.  

For the remaining seven candidate white dwarfs with matching radio sources, spectra were obtained with SOAR/Goodman \citep{goodman} over several nights between 2022 December 2 and 2023 May 16. These observations all used the 400 l mm$^{-1}$ grating and a 1.2\arcsec\ slit, covering an approximate usable wavelength range 4000--7850~\AA\ at a mean full-width at half-maximum (FWHM) resolution of $\sim 6.6$~\AA. A single initial spectrum of length 1200--1800~s was obtained for each candidate. Using the same setup, three follow-up spectra of WDJ121604.90-281909.67 were also taken, each on a different night over approximately one month. Finally, we also observed WDJ204259.71+152108.06 using two additional setups, comprising: (i) a single spectrum taken on 2023 June 16 with the 400~l~mm$^{-1}$ grating with redder wavelength coverage (4800--8800~\AA), and (ii) 61 spectra at medium resolution, using a 1200~lines~mm$^{-1}$ grating covering 7770--8870~\AA\ at a mean FWHM resolution of $\sim 2.2$~\AA, obtained from 2023 Jun 15 to Oct 25. All SOAR/Goodman spectra were reduced and optimally extracted in the standard manner using IRAF \citep{Tody1986}. While all spectra were wavelength calibrated with arc lamp spectra taken immediately adjacent to the object spectra, we also made small zeropoint corrections to the wavelength scale using the telluric A band (for 400~l~mm$^{-1}$ spectra) or narrow water features around 8228~\AA\ (for the 1200~l~mm$^{-1}$ spectra). Additionally, two spectra of WDJ204259.71+152108.06 were obtained with the Kast double spectrograph at the 3-m Shane Telescope on 26 July 2022. We only use the blue spectrum here, which was obtained with the 600/4310 grism (covering 3300-5520~\AA). A 2'' slit was used, and conditions were clear, with 1.5—2'' seeing. The data were reduced with PypeIt \citep{Prochaska2020}, using standard bias subtraction and flat fielding techniques. Internal He, Hg, Cd, Ne lamps were used to obtain a wavelength solution.

\begin{figure}
    \includegraphics[width=\columnwidth]{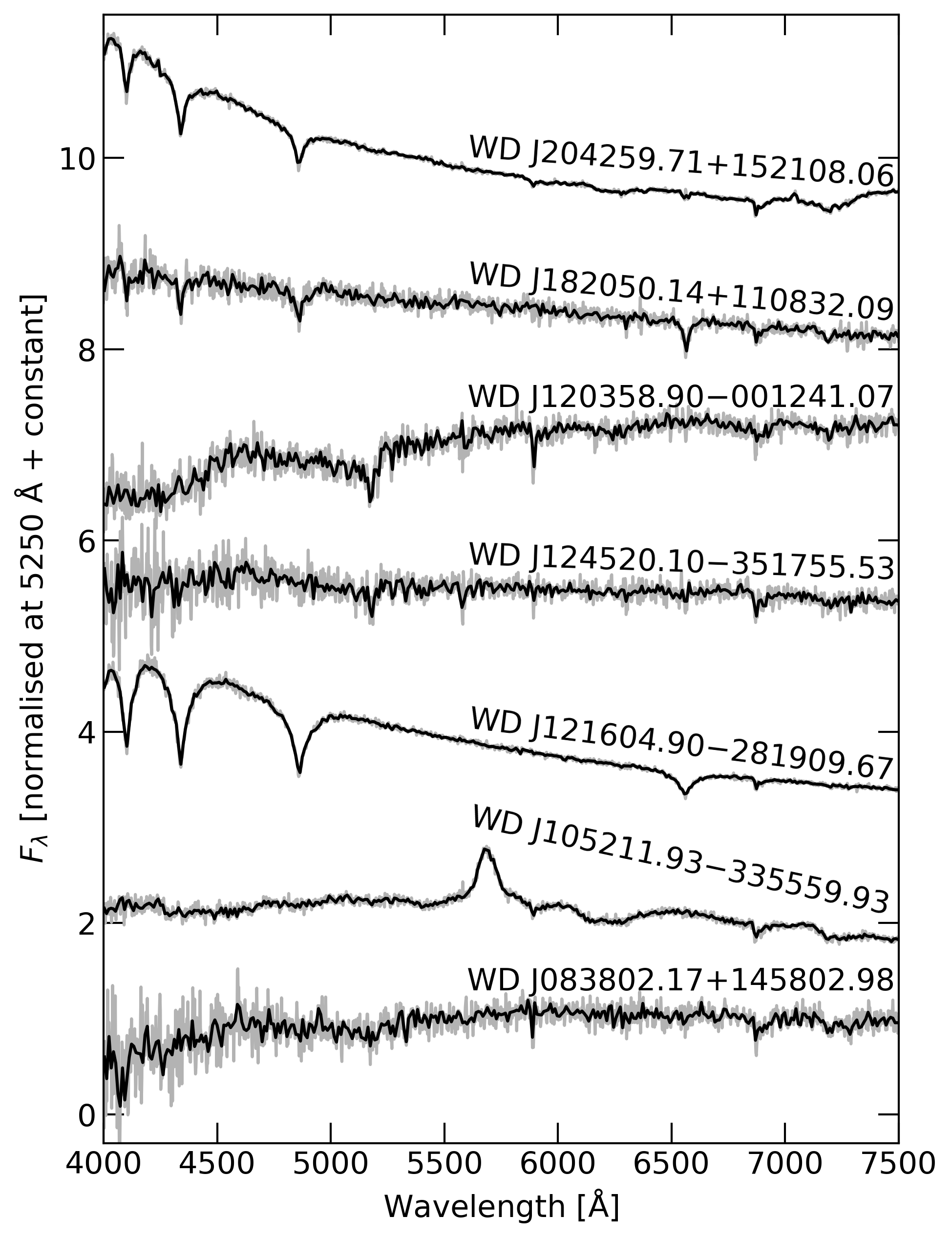}
    \caption{SOAR spectra for the seven white dwarf candidates with matching radio sources. Fluxes have been normalised to 1.0 at 5250~\AA\ and displaced vertically by adding a constant for clarity. The grey line shows the observed spectrum, with the black line showing a five-point average.}
    \label{fig:spectra}
\end{figure}

The low-resolution classification spectra are shown in Figure~\ref{fig:spectra}. Inspection of the obtained spectra immediately reveals that at least four out of the seven objects are not white dwarfs. The broad emission line shown by WD~J105211.93-335559.93 points at a quasar nature. Like the other quasar identified in this sample, it also has a significant parallax of $2.52\pm0.44$~mas. These two objects exemplify how parallax precision does not imply accuracy. As our cross-matching selects on radio emission, which is an uncommon characteristic for white dwarfs, it is not surprising that our selection predominantly reveals contaminants amongst the white dwarf candidates.

The candidates WD~J083802.17+145802.98, WD~J124520.10-351755.53 and WD~J120358.90-001241.07 show spectra typical of late-type main sequence stars, with spectral types between F and K. They all have low white dwarf probabilities and lie around $G_{BP} - G_{RP} = 1$, in the colour region of the main sequence turn-off (Figure \ref{fig:cmd}). The large number of stars in this region facilitates the occurrence of statistical extremes for which the true parallax value is many standard deviations away from the reported value. These three objects are likely examples of such extremes. As we focus here only on white dwarfs, we make no further attempt to determine whether the radio emission indeed matches these stars, though it is worth noting that WDJ120358.90-001241.07 is the one object with no identified background contamination but for which the separation between radio and optical is larger than 1". That does not necessarily mean that these sources are not genuine radio emitters, as radio emission is sometimes detected in late-type main sequence stars due to gyrosynchroton emission from mildly relativistic electrons \citep[e.g.,][]{Gudel2002}.

The follow-up spectra leave three out of our original candidates showing spectra consistent with white dwarfs. The two candidates with the highest $P_\mathrm{WD}$ are confirmed as white dwarfs, with WDJ121604.90-281909.67 showing a spectrum typical for a hydrogen-dominated (DA) type white dwarf, and WDJ182050.14+110832.09 displaying a similar spectrum, but with a cooler temperature resulting in weaker lines and less steep continuum. WDJ204259.71+152108.06 was observed in morning twilight, with considerable background affecting the red part of the spectrum, but features observed suggest a composite spectrum, dominated by a DA white dwarf in the blue with contribution from a late-type companion (likely a M-dwarf) in the red.

\begin{figure}
    \includegraphics[width=\columnwidth]{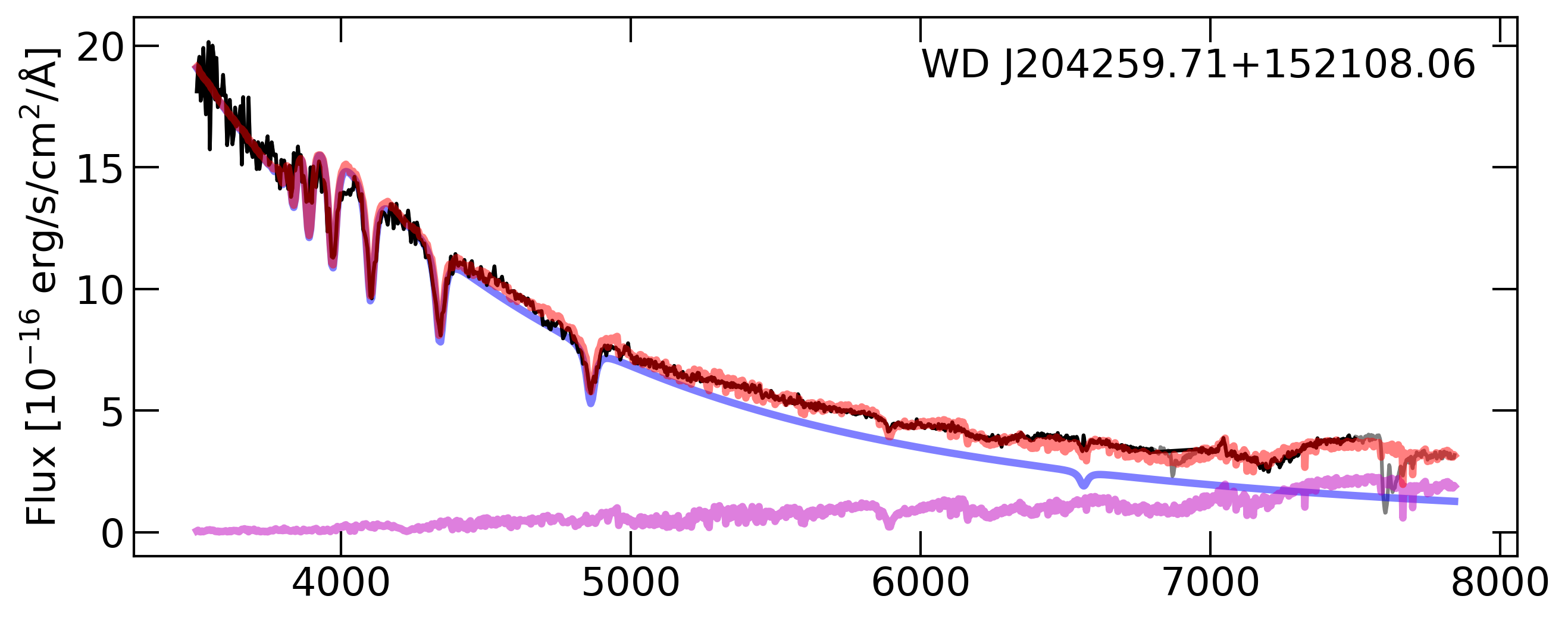}
    \includegraphics[width=\columnwidth]{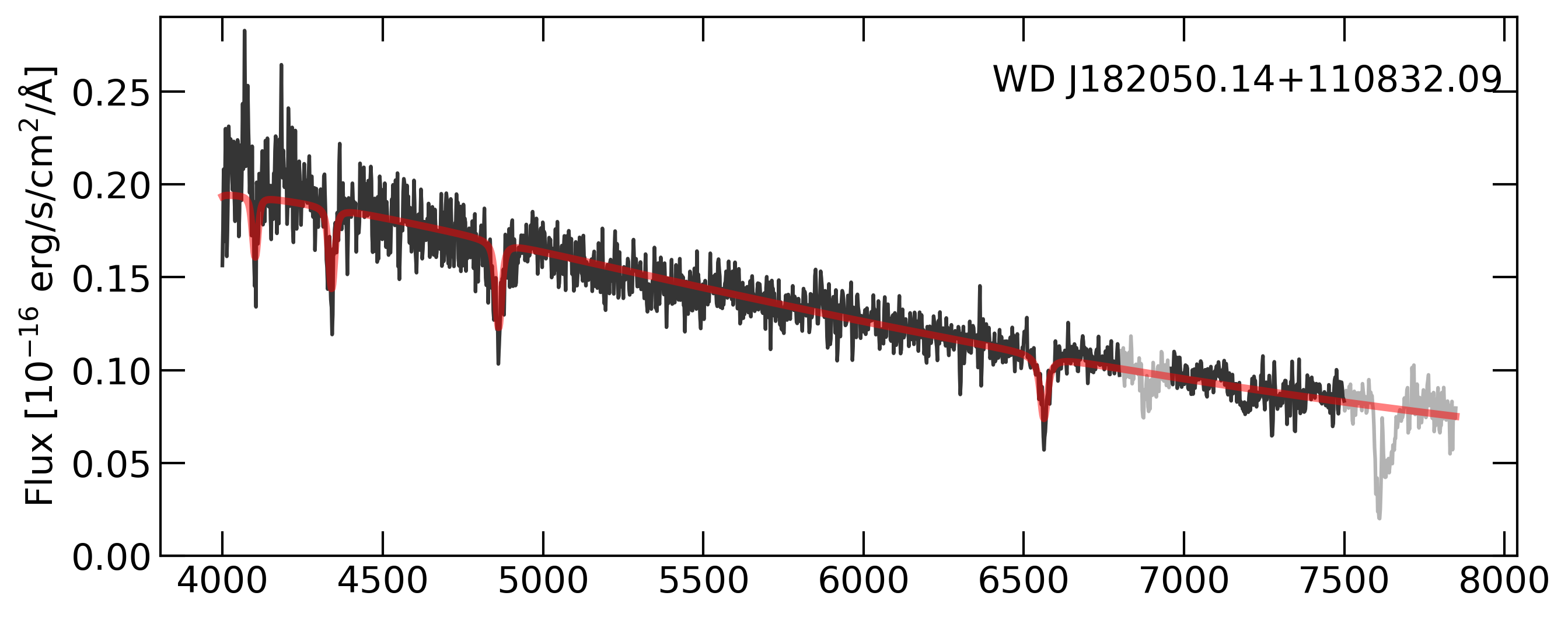}
    \includegraphics[width=\columnwidth]{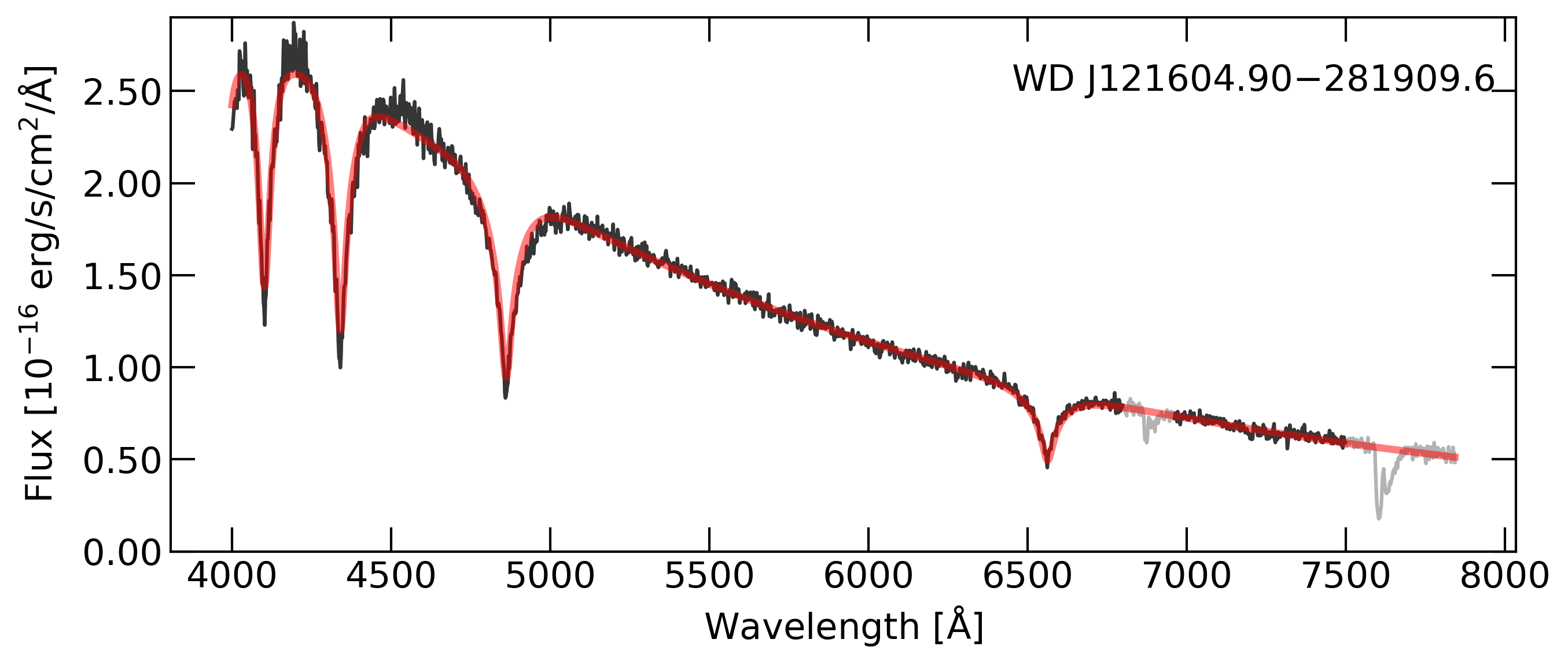}
    \caption{Spectral fits for WD~J121604.90-281909.67 (bottom), WD~J182050.14+110832.09 (middle), and WD~J204259.71+152108.06 (top). The observed spectra are shown in black, with regions masked in the fitting (due to telluric contributions) shown in grey. The best-fit model is shown in red. For WDJ204259.71+152108.06, we show the best fit white dwarf model in blue and the M-dwarf model in magenta.}
    \label{fig:specfit}
\end{figure}

We fit the spectra of these three white dwarfs using models calculated by \citet{Koester2010}, with pure-hydrogen atmospheres and convection implemented using the mixing-length approximation with the mixing-length scale $\alpha$ set to 0.8. The grid covers $5\,000 < T_\mathrm{eff} < 80\,000$~K and $6.5 < \log~g < 9.5$. We calculated the $\chi^2$ between the observed spectra and every model in the grid, first degrading the models to the resolution of the observed spectra. Spectra and models were normalised by a constant free parameter, preserving the slope but not the absolute flux given potential slit losses. In all cases an absolute minimum was found, and we adopt the best solution as the one given by the global minimum. Uncertainties were obtained from the 68\% percentile of the $\chi^2$ distribution. The spectra observed were corrected for extinction prior to fitting using $E(B-V)$ values from a 3D extinction map \citep{Capitanio2017} and the {\it Gaia} geometric distance \citep{Bailer-Jones+2021}. For WD~J204259.71+152108.06, we combine the SOAR spectrum with the KAST spectrograph blue channel spectra (the red channel spectra were unusable due to a short exposure length) by using the KAST spectrum for wavelengths below 5000~\AA\ and the SOAR spectrum above that. No renormalisation was required, giving confidence that the absolute flux calibration is reliable. We then simultaneously fit the white dwarf and the M-dwarf, using BT-Settl models \citep{Allard2013} for the M-dwarf (with $\log~g = 5$, solar metallicity, and no $\alpha$-enhancement). The resulting fits are shown in Figure~\ref{fig:specfit}.

{\bf WD~J121604.90-281909.67} is found to show $T_\mathrm{eff} = 11\,000\pm250$~K and $\log~g = 8.65\pm0.35$. Using a mass-radius relationship from carbon-oxygen core cooling models{\footnote{\url{https://www.astro.umontreal.ca/~bergeron/CoolingModels/}}} \citep{Bedard2020}, these parameters indicate a mass of $0.94\pm0.15$~M$_{\sun}$. All of the obtained values agree within $2\sigma$ with the values reported by \citet{GentileFusillo+21} for a hydrogen-atmosphere spectral energy distribution fit ($11\,500\pm800$~K, $8.20\pm0.17$, $0.73\pm0.11$~M$_{\sun}$).

{\bf WD~J182050.14+110832.09}'s spectrum was fitted with $T_\mathrm{eff} = 6850\pm160$~K and $\log~g = 7.3^{+0.8}_{-1.2}$. At such low temperature, the lines become less sensitive to $\log~g$, which results in high uncertainty. The mass is estimated to be $0.51\pm0.25$~M$_{\sun}$ from carbon-oxygen cooling models \citep{Bedard2020}. Like for WD~J121604.90-281909.67, the derived parameters are consistent with those obtained by \citet{GentileFusillo+21} ($6300\pm700$~K, $7.6\pm0.5$, $0.40\pm0.21$~M$_{\sun}$). We note that {\it Gaia}'s astrometric solution for WD~J182050.14+110832.09 shows some excess noise ({\tt astrometric\_excess\_noise} =1.3~mas), but at $G = 20$~mag that is not unexpected.

{\bf WD~J204259.71+152108.06} was fitted with $T_\mathrm{eff} = 30\,000^{+7000}_{-5000}$~K and $\log~g = 6.5^{+1.5}_{-1.0}$, implying a mass of $0.5^{+0.2}_{-0.1}$~M$_{\sun}$. The M-dwarf was best fitted with $T_\mathrm{eff} = 3\,800\pm200$~K, corresponding to spectral type $\sim$M2.5. The large uncertainties are due to the different combinations of white dwarf and M-dwarf parameters that result in similar $\chi^2$. For this system, given the evidence that slit losses were minimal, the normalisation was set by the radii of each star and the distance of the system to Earth, which were left to vary freely. The best fit was obtained with $R = 0.015^{+0.02}_{-0.01}$ and $R = 0.24^{+0.04}_{-0.01}$~R$_{\sun}$ for the white dwarf and M-dwarf, respectively, and distance $284^{+13}_{-3}$~pc. \citet{GentileFusillo+21} provided no parameter estimates for this system, because of its low white dwarf probability that can be attributed to the companion contribution moving it away from the white dwarf cooling track. The presence of a clear companion makes WD~J204259.71+152108.06 particularly interesting as a radio source, as the acceleration of electrons due to interaction between the stars could lead to genuine radio emission. Of the three targets confirmed as white dwarfs, this one has the lowest astrometric  separation between white dwarf position and radio source.
 
\subsection{Time-series photometry}
\label{sec:phot}

White dwarf systems known to emit in radio are all also variable in the optical, as is the case for magnetic propellers \citep[e.g.][]{Patterson1979}, binary white dwarf pulsars \citep[e.g.][]{Marsh2016}, and cataclysmic variables in general \citep[e.g.][]{Scaringi2014}. To identify whether the three sources confirmed as white dwarfs could be associated with one of these known types of radio emitters, we searched for variability in their light curves from the Asteroid Terrestrial-impact Last Alert System \citep[ATLAS,][]{atlas}, the Zwicky Transient Facility \citep[ZTF,][]{ztf}, and the Catalina Surveys \citep{catalina}. ATLAS measurements are available in two bands, cyan ($c$) and orange ($o$); ZTF provide data in $g, r, i$, and Catalina measurements are in the $V$ band. The targets were also followed up with the fast-speed photometers ULTRACAM \citep{ultracam}, mounted on the 3.5-m European Southern Observatory (ESO) New Technology Telescope (NTT), and ULTRASPEC \citep{ultraspec}, mounted on the 2.4-m Thai National Telescope (TNT), as fast variability such as the white dwarf spin can be missed by the cadence of surveys.

We first searched for periodic variability by calculating Lomb-Scargle periodograms \citep{Lomb1976, Scargle1982} for each band separately. We probed periods between one minute and five days, spanning a typical range of white dwarf spin periods \citep{Hermes2017} and orbital periods where binary effects from proximity would be detectable in photometric data. To identify stochastic variability, our approach was to compare the distribution of the number of standard deviations ($n \sigma$) from the mean ($\mu$) with the maximum expected $n$ given the number of observations, assuming that the measurements would follow a Gaussian distribution for a non-variable star. The probability that a measurement of a Gaussian variable lies in the range between $\mu - n\sigma$ and $\mu + n\sigma$ is given by
\begin{equation}
   P_n = \mathrm{erf}\left( \frac{n}{\sqrt{2}} \right),
\end{equation}
where erf is the error function. Hence for a number $N_\mathrm{obs}$ of observations, there is a given $n$ for which
\begin{equation}
    N_\mathrm{obs} \times \mathrm{erf}\left( \frac{n}{\sqrt{2}} \right) < 1,
\end{equation}
i.e. all observations are within the expected range assuming that any scattering is due to only random noise. We solved for $n$ in each case and compared it with the observed $n$ distribution. In the ideal case, no $i$ measurement should have $n_i > n$, but does not account for systematic uncertainties. In an attempt to consider systematics, we impose a variability threshold of 10 per cent of the $n$ distribution being above the expected $n$, i.e. we assume that up to 10 per cent of measurements can be adversely affected by systematics.

\begin{figure*}
    \includegraphics[width=\textwidth]{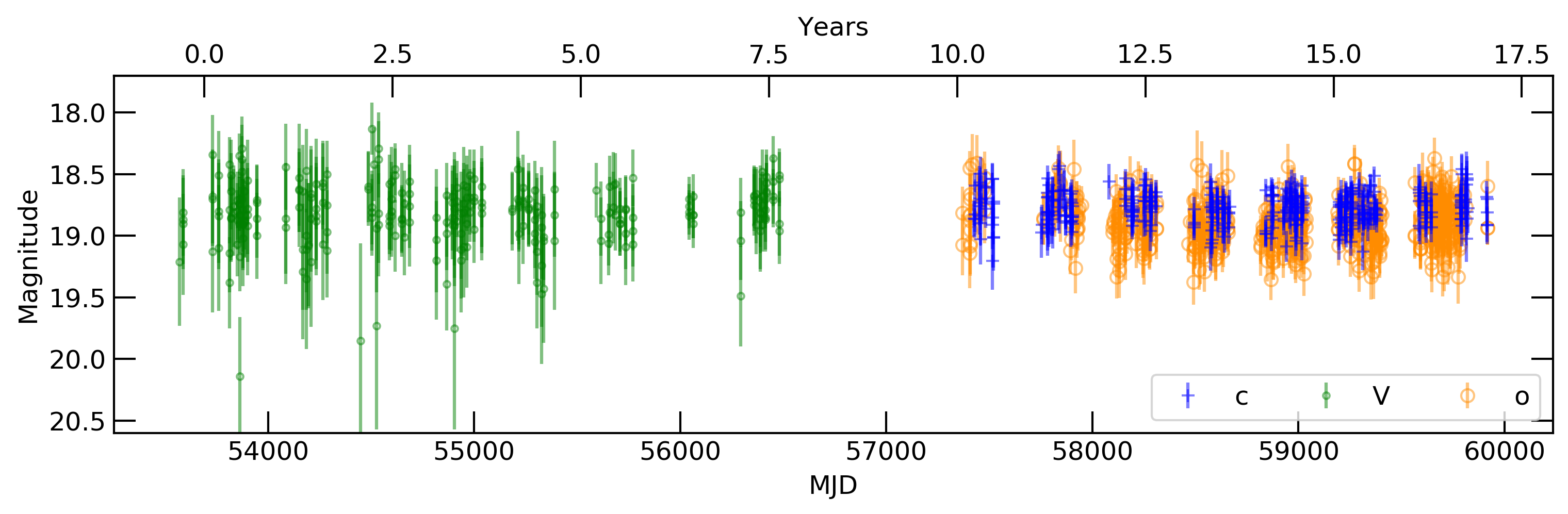}
    \caption{ATLAS (blue crosses for $c$ and orange squares for $o$) and Catalina (green circles) light curves for WD~J121604.90-281909.67, spanning over 17 years.}
    \label{fig:j1216phot}
\end{figure*}

{\bf WD~J121604.90-281909.67} had available photometry from ATLAS (spanning 2015--2022) and Catalina (spanning 2005--2013), which is shown in Fig.~\ref{fig:j1216phot}. A Lomb-scargle periodogram for each dataset is shown in Fig.~\ref{fig:j1216ft}. There are no clear periods rising above the noise, with the strongest peaks being attributed to one-day aliases resulting from sampling. Periodic variability can be ruled out over the range of probed periods (1 min to 5 days) to about 50~mmag. We also identify no stochastic variability in these light curves. The Catalina light curve has 0.4 per cent of measurements beyond the expected $n$ of 2.9 $\sigma$ from the mean. For ATLAS, 1.6 per cent of $o$ measurements and $0.5$ per cent of $c$ are beyond the expected $n$. Therefore there is no indication of any variability in the archival light curves.

To probe for short-term or low-level variability, we observed WD~J121604.90-281909.67 with ULTRASPEC on 2022 December 19 and 2023 January 25. In both cases we employed the $g_s$ filter; this filter has approximately the same cut-off range as the traditional $g$ filter but with higher throughput (indicated by the subscript $s$ for ``super"). Cadence was set as 10.05 and 14.98~s and the length of exposure as 109 and 99~min in the first and second nights, respectively. Deadtime is negligible thanks to frame-transfer capabilities. The light curve and Fourier transforms are shown in Fig.~\ref{fig:j1216uspec}. There are no significant peaks in the periodogram and periodic variability up to 3 hours can be ruled out down to an amplitude of less than 1 per cent. No indication of stochastic variability is found either.

\begin{figure*}
    \includegraphics[width=\textwidth]{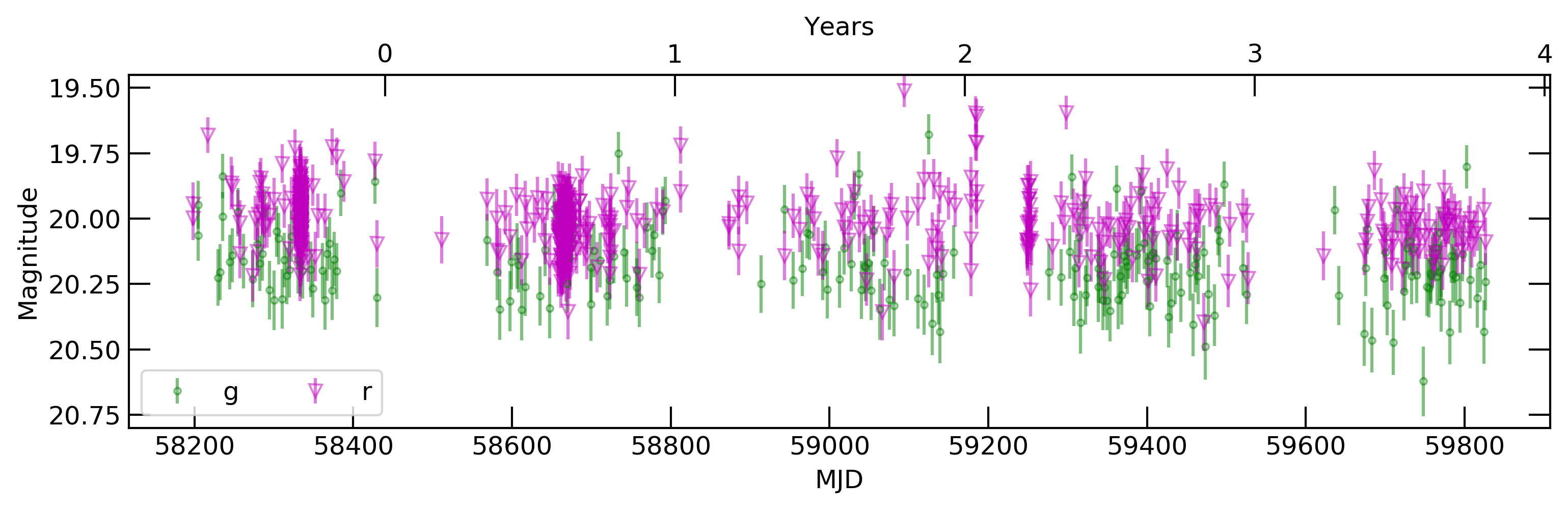}
    \caption{ZTF light curves for WD~J182050.14+110832.09, with $g$ shown as green circles and $r$ as magenta triangles.}
    \label{fig:j1820phot}
\end{figure*}

{\bf WD~J182050.14+110832.09} was observed by ZTF in the $g$ and $r$ bands, as shown in Fig.~\ref{fig:j1820phot}. Publicly available data span 2018--2021. The light curve shows considerable scatter, but at 20th magnitude this target approaches the detection limit of ZTF. A Lomb-Scargle periodogram reveals no periodic variability between 1 min and 5 days down to $\approx 50$~mmag (Fig.~\ref{fig:j1820ft}). The only strong peaks are due to one-day aliasing. 6.6 and 2.3 per cent of $g$ and $r$ measurements, respectively, are above the expected maximum $n$, below our variability threshold.

We subsequently observed WD~J182050.14+110832.09 with ULTRACAM on 2023 April 23 for an hour. ULTRACAM's beam splitter allows simultaneous observations in three filters. Filters $u_s$, $g_s$, and $r_s$ were employed. The cadence was 10.06~s for $g_s$ and $r_s$, and $30.18$~s in $u_s$. Like ULTRASPEC, ULTRACAM also has negligible deadtime. The light curve and periodogram are shown in Fig.~\ref{fig:j1820ucam}. No significant peaks are seen in the periodogram, ruling out periods up to two hours down to 2 ($g_s, r_s$) or 10 ($u_s$) per cent amplitude (the one peak seen in the $u_s$ periodogram is due to a cadence alias). There is also no sign of stochastic variability, with less than 1 per cent of the measurements being above the expected maximum $n$ in all cases. 

\begin{figure*}
    \includegraphics[width=\textwidth]{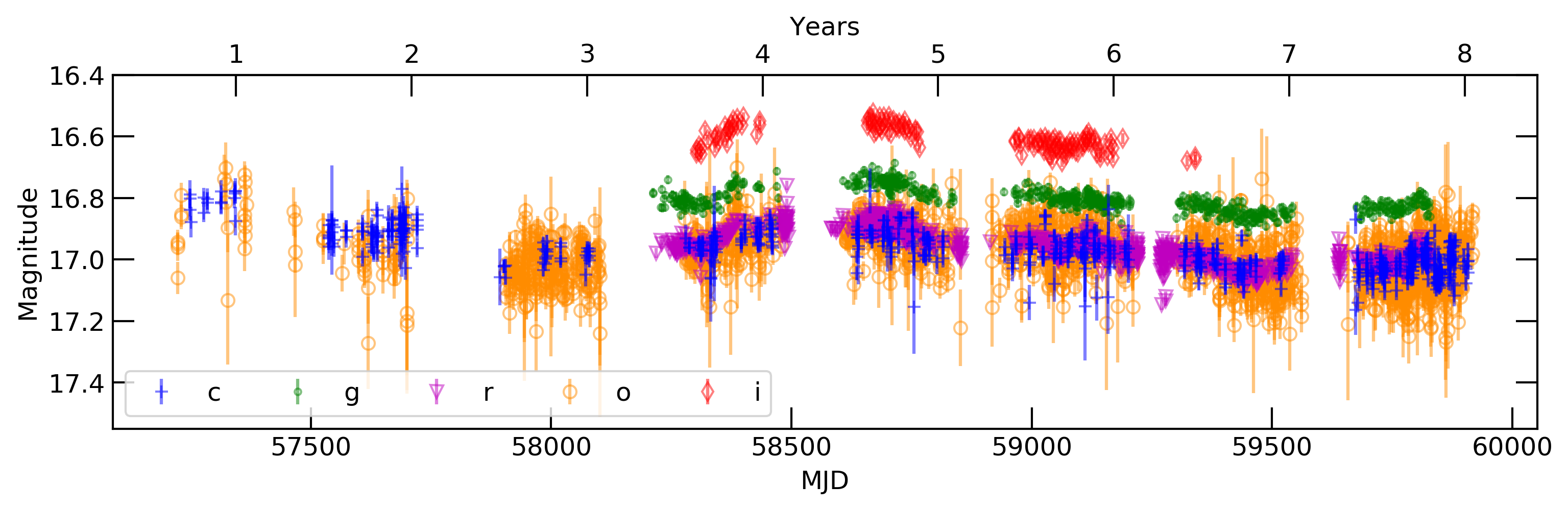}
    \caption{ATLAS ($c$ as blue crosses and $o$ as orange circles) and ZTF light curves ($g$ as green dots, $r$ as magenta squares, and $i$ as red diamonds) for WD~J204259.71+152108.06. Note the increase in brightness visible just before MJD = 58500, and subsequent decrease around a year after.}
    \label{fig:j2042phot}
\end{figure*}

{\bf WD~J204259.71+152108.06} has light curves in both ATLAS (2015--2022) and ZTF (2018--2022), which are shown in Fig.~\ref{fig:j2042phot}. Visual inspection alone already suggests some long-term variability at the level of $\approx 0.2$~mag. We indeed find evidence for stochastic variability in our analysis, which showed that 18, 25, 35, 13, and 29 per cent of measurements are beyond the expected $n$ for $c$, $g$, $r$, $o$, and $i$, respectively, above our variability threshold in all cases. The variability does not seem to be periodic in nature, as shown in the periodograms in Fig.~\ref{fig:j2042ft}. Many peaks are present, but can be attributed to one-day aliases.

We observed WD~J204259.71+152108.06 with ULTRACAM on two nights to probe for short-term variability, in both nights using $u_s$, $g_s$ and $r_s$ filters. First we observed it for 1.5~hours on 2023 September 03, and then for an hour on 2023 September 14. The cadence was 4.04~s on the first night and 2.84~s on the second night in $g_s$ and $r_s$ bands, and three times that in the $u_s$ band. Figure~\ref{fig:j2042ucam} shows the light curves and periodograms for the data. We find no indication of periodic variability down to an amplitude of $\approx 0.3$ per cent. Similarly, in these short runs there is also no evidence for aperiodic variability. We conclude that WD~J204259.71+152108.06 may show long-term variability that is likely aperiodic (or at least with a period longer than our 8-year baseline), as indicated by ATLAS and ZTF, but no detectable periodic or short-term variability.

\subsection{High-resolution radio imaging}
\label{sec:highrad}

To further test the association of our three spectroscopically confirmed white dwarfs with the VLASS radio sources, we sought higher resolution radio imaging which could yield more precise astrometry. The below-described radio data were all calibrated and imaged using standard routines in the Astronomical Image Processing System (AIPS; \citealt{Greisen03})\footnote{http://www.aips.nrao.edu/index.shtml}. All imaging was carried out using a Briggs robust value of 1, and all equatorial coordinates are on the ICRS.

To explore the radio sky around {\bf WD~J121604.90-281909.67} at higher resolution, we applied for director's discretionary time on the VLA (program code 23A-389; PI L.\ Chomiuk) while it was in its B configuration. Data were obtained over a half hour on 2023 April 28 in the X band (8--12 GHz), yielding 13 minutes on source. We chose a different band than VLASS observations to probe the spectral index of the source. To image the field, the 4\,GHz of bandwidth was split in two to yield two images centered at 8.9 and 11.0 GHz. The FWHM of the synthesized beam in the 8.9 GHz image is $1.9'' \times 0.7''$ (PA $= 26^{\circ}$) and 1$\sigma$ noise of 11 $\mu$Jy beam$^{-1}$. At 11.0 GHz, the beam FWHM is $1.5'' \times 0.6''$ (PA $= 26^{\circ}$) with a 1$\sigma$ noise of 11 $\mu$Jy beam$^{-1}$. In both images, the radio source is consistent with a point source at a position of RA = 12h16m04.7963s, Dec = $-28^{\circ}19'09.738''$; at 8.9 GHz it has a flux density of $0.890 \pm 0.013$ mJy, while at 11.0 GHz the flux density is $0.708\pm0.012$ mJy. Comparing these measurements with the VLASS flux densities implies a relatively steep radio spectrum ($\alpha = -0.9$ where $S_\nu \propto \nu^{\alpha}$), as expected for optically thin synchrotron emission.
Taking the \emph{Gaia} astrometric parameters and advancing them to the epoch of this radio observation, WD~J121604.90-281909.67 would have a position of RA = 12h16m04.8437s, Dec = $-28^{\circ}19'09.786''$, separated from the radio source by 0.63". The complex gain calibrator used in our VLA observation (1209$-$241) has a position accuracy code of A, implying that its position is known to $<0.002''$. A positional accuracy of about a tenth of the synthesised beam's FWHM ($< 0.2 ''$) should therefore be attainable at the field's phase centre where WD~J121604.90-281909.67 is located\footnote{\url{https://science.nrao.edu/facilities/vla/docs/manuals/oss/performance/positional-accuracy}}. We conclude that the separation between the white dwarf and the radio source is significantly larger than the uncertainty on the position of the radio source or the white dwarf, implying that the radio source is unlikely to be associated with the white dwarf (and probably attributable to an optically faint background galaxy).

To further investigate {\bf WD~J182050.14+110832.09}, we downloaded archival data from the pre-upgrade Very Large Array, obtained in the X band (8.5 GHz) using the VLA's A configuration under program code AM623 (PI S.\ Myers). The data were obtained on 1999 Aug 17.3, with just 43 seconds integrated on source and the radio source located very near the phase centre (1.4 arcsec away). The resulting image has a synthesised beam FWHM of $0.38'' \times 0.25''$ (PA $= 57^{\circ}$) and 1$\sigma$ noise of 0.23 mJy beam$^{-1}$. The radio source is consistent with a point source and has a flux density of 24.7 mJy, significantly brighter than the VLASS flux measurements (Table \ref{tab:wd_vlass}). This may imply that the radio spectrum is inverted (increasing to high frequency; $\alpha \approx 0.5$) as expected for emission that is partially optically thick, but it may also be explained with variability (there is significant variability measured between VLASS Epochs 1 and 2; Table \ref{tab:wd_vlass}). The radio source is located at RA = 18h20m50.1090s, Dec = $+11^{\circ}08'32.423''$. Accounting for the proper motion measured by \emph{Gaia}, 
WD~J182050.14+110832.09 would have a position of RA = 18h20m50.1384s, Dec = $+11^{\circ}08'32.070"$ at the epoch of the radio observations. Therefore, the separation between the white dwarf and the radio source is $0.56''$. The complex gain calibrator used in this VLA observation (1824$+$114) has a position accuracy code of B, implying that its position is known to $0.002-0.01''$. Hence, the uncertainty on the position is probably best given as $\sim$0.1 FWHM of the synthesised beam ($\sim0.04''$). As the separation between the white dwarf and radio source is significantly larger than the uncertainty on either position, we conclude that the radio source is probably not associated with the white dwarf and is a rather a chance near-coincidence.

Higher resolution observations of {\bf WD~J204259.71+152108.06} were found in the NRAO archive, observed with the A configuration of the Jansky VLA on 2019 Sep 12.0, using the C band (4--8 GHz) under program code SB072. A field including WD~J204259.71+152108.06 was observed in a single 103-s long scan, with the white dwarf position offset from the phase centre by $3.4'$ (this is relatively far out in the C-band primary beam, whose half-power point is $4.5'$ from the phase centre at 5 GHz and $3.2'$ at 7.1~Hz). To image, we split the 4~GHz of bandwidth in two halves to yield two images centred at 5.0 and 7.1~GHz. The images were created with faceting to correct for wide-field effects, and were primary-beam corrected before measuring the source's flux density. At 5~GHz, the resulting synthesised beam has FWHM of $0.42'' \times 0.37''$ (PA $= -12^{\circ}$) and 1$\sigma$ noise of 29~$\mu$Jy beam$^{-1}$, while at 7.1~GHz the point spread function's dimensions are $0.30'' \times 0.26''$ (PA $= -14^{\circ}$) with 1$\sigma$ noise of 32 $\mu$Jy beam$^{-1}$. The radio source is detected at high significance and is consistent with a point source in both images at a position of RA = 20h42m59.7254s, Dec = $+15^{\circ}21'08.293''$. At 5.0~GHz, we measure a flux density of $9.25\pm0.05$~mJy, and at 7.1~GHz, the flux density is $9.21\pm0.08$~mJy. When compared with the VLASS flux densities measured at 3 GHz, the radio spectrum appears to be marginally steep to flat ($\alpha \approx -0.3$). Accounting for the \emph{Gaia} proper motion, 
WD~J204259.71+152108.06 would have a position of RA = 20h42m59.731s, Dec = $+15^{\circ}21'08.138''$ at the time of these radio observations, offset from the position of the radio source by $0.18''$. 

This is the smallest optical--radio separation of the sources considered in this paper, but does it imply that the radio source and white dwarf are associated? The \emph{Gaia} parameter RUWE is very large for this source (7.3), implying a poor astrometric solution, although it is not simple to quantify the resulting positional uncertainty. The position of the complex gain calibrator used for the radio position 2031+123 is accurate to $<0.002''$. As previously discussed, while a positional accuracy of about a tenth of the synthesised beam's FWHM is possible, this accuracy is only reached at the field's phase centre. In these high-resolution VLA observations, the white dwarf is located rather far out in the primary beam and errors on position grow with distance from the phase centre\footnote{\url{https://science.nrao.edu/facilities/vla/docs/manuals/oss/performance/positional-accuracy}}. Given the significant but poorly defined uncertainties on both the \emph{Gaia} and radio astrometry of this source, we consider it possible that the white dwarf and the radio source are associated.

To assess the variability of the radio source potentially associated with WD~J204259.71+152108.06, we downloaded lower-resolution VLA observations obtained in the B configuration and the L band (1--2 GHz) under program code 20A-439 (PI S.\ Bruzewski). The observation was obtained on 2020 Oct 9 and yielded 60 minutes on source at the same pointing centre as the C band observations (but as the L-band primary beam is larger --- FWHM = $30'$ at 1.5 GHz --- this location is less far out in the primary beam). For imaging, we split the bandwidth in two halves, yielding images at 1.36 GHz (synthesised beam FWHM $\approx4.7''$) and 1.83 GHz (synthesised beam FWHM $\approx3.5''$). The  observation-averaged flux density is $12.41\pm0.03$~mJy at 1.36~GHz and $12.56\pm0.02$~mJy at 1.83 GHz. To look for variability over this observation, we separately imaged the five 12-minute-long observation scans. We see no evidence for significant variations in the flux density. For comparison, the white dwarf pulsar AR~Sco shows factor of $\sim$2 variability over the course of its 3.56\,h orbital period \citep{Stanway+18}, which would have been easily detectable in this radio observation.

Because AR Sco's radio emission is highly circularly polarised during some phases of its orbit \citep[up to 30 per cent][]{Stanway+18}, we investigated Stokes $V$ images of WD~J204259.71+152108.06. The 7.1 GHz image yields a non-detection in Stokes $V$, implying a 3$\sigma$ upper limit on the circular polarisation fraction of $<2.6\%$. At 5.0 GHz, there is a low-significance (5$\sigma$) Stokes $V$ positive signal at the position of the Stokes $I$ source, implying circular polarisation at the level of $2.3\pm0.5\%$. We however caution that this estimate is likely affected by circular beam squint, due to the fact that we are observing relatively far out in the primary beam, which can lead to uncertainties of $\sim$few percent\footnote{\url{https://science.nrao.edu/facilities/vla/docs/manuals/obsguide/modes/pol}}.
The lower-frequency observation yielded detections of negative (right-handed) Stokes V signal, implying circular polarisation fractions of $-1.8\pm0.1\%$ at 1.36 GHz and $-1.9\pm0.1\%$ at 1.83 GHz. 

\subsection{Time-series spectroscopy}

For WD~J204259.71+152108.06, which showed evidence of a cool M-dwarf companion star and a potentially consistent astrometric position between the white dwarf and the radio emission, we obtained time-series spectroscopy with SOAR at moderate resolution, encompassing 61 spectra taken on 9 nights over a time span of 132~d. We derived barycentric radial velocities for these spectra through cross-correlation with a synthetic M dwarf template with $T_{\rm eff} = 3500$ K in the region of the \ion{Na}{1} doublet at 8190 \AA.

\begin{figure}
\includegraphics[width=\columnwidth]{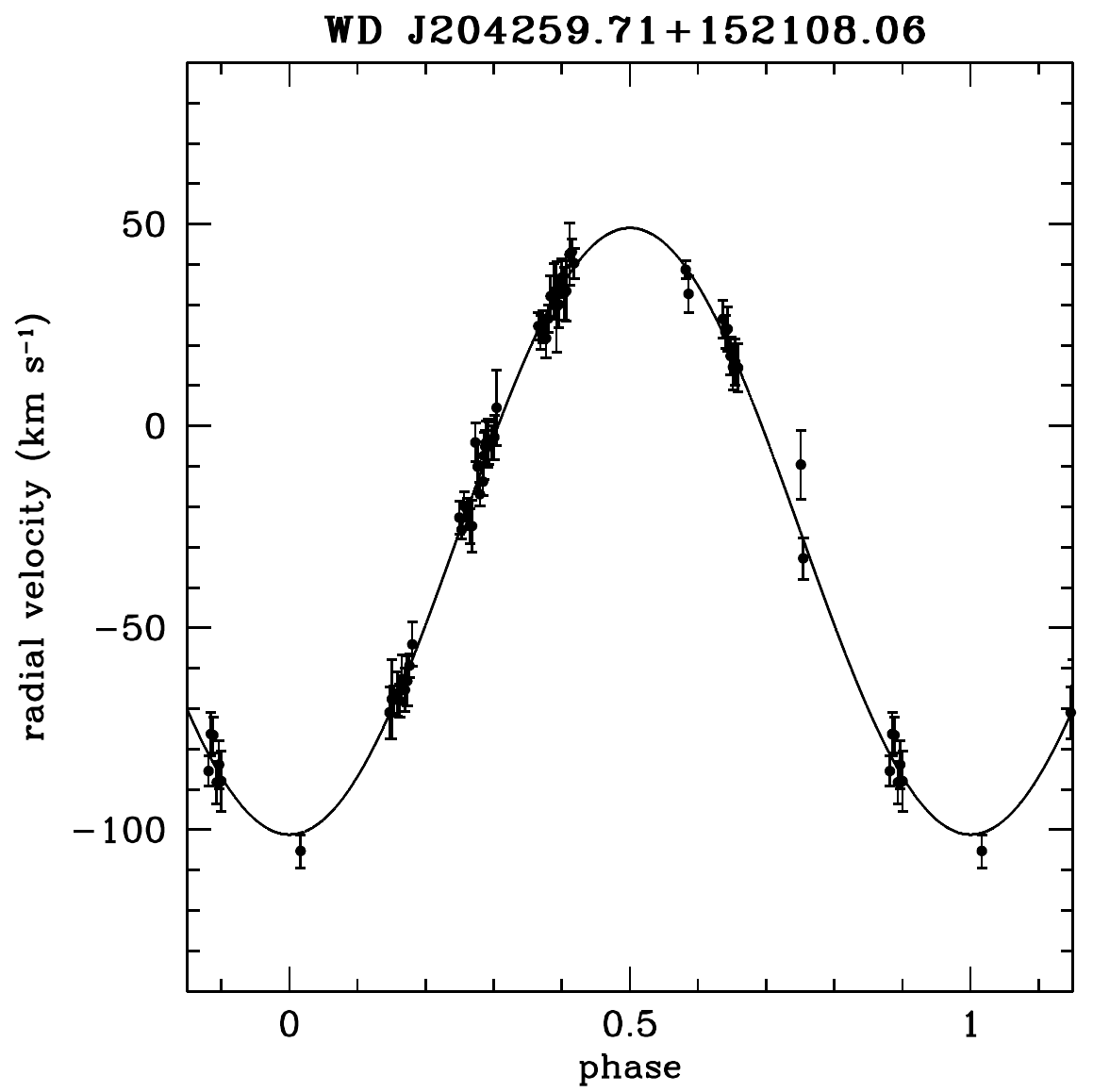}
    \caption{Radial velocities for the M dwarf companion to WD~J204259.71+152108.06, folded on the best-fit orbital period of 4.15433 d. The best-fit circular Keplerian model is overplotted.}
    \label{fig:rv_fig}
\end{figure}

We fit these velocities using the custom
Markov Chain Monte Carlo sampler \texttt{TheJoker} \citep{Price-Whelan2017}, 
finding an excellent fit (r.m.s. = 4.2 km s$^{-1}$) to a circular model with binary period $P = 4.15433(80)$ d, BJD time of ascending node of the white dwarf $T_{0} = 60203.4397(76)$ d, semi-amplitude $K_{2} = 75.1 \pm 1.0$ km s$^{-1}$, and systemic velocity = $-26.1\pm0.7$ km s$^{-1}$, with all uncertanties listed as $1\sigma$ estimates. This fit is plotted with the data in Figure~\ref{fig:rv_fig}. No other period has posterior support, and all parameters are well-constrained. Allowing a non-zero eccentricity does not improve the fit quality and the posterior of the eccentricity parameter is strongly peaked at zero; hence, we retain the circular fit as our best model.

The binary mass function $f(M)$,
\begin{equation}
f(M) = \frac{P K_{2}^{3}}{2 \pi G} = \frac{M_1 \, (\textrm{sin}\, i)^{3}}{(1+q)^{2}}
\end{equation}
for mass ratio $q=M_{2}/M_{1}$ and inclination $i$, has a measured value of 
$f(M) = 0.182(8) M_{\odot}$. Unfortunately, this value provides not much more than a consistency check: 
for a $\sim 0.25 M_{\odot}$ M-dwarf secondary and a $\sim 0.5 M_{\odot}$ white dwarf primary, the inferred inclination is around $69^{\circ}$, but a broad range of inclinations and secondary masses are consistent with the mass function and the uncertain white dwarf mass of $0.5^{+0.2}_{-0.1} M_{\odot}$, excepting perhaps face-on ($\lesssim 40^{\circ}$) orbits. Given the estimated M dwarf radius of $\sim 0.24\, R_{\odot}$, eclipses of the white dwarf would only be expected for exceptionally edge-on ($\gtrsim 89^{\circ}$) orbits,
so the observed lack of eclipses does not meaningfully constrain the inclination. In fact, the lack of photometric signature on the orbital period (Section~\ref{sec:phot}) is unsurprising, as no significant irradiation or tidal effects are expected.

Overall, the optical spectroscopy for WD~J204259.71+152108.06 is fully consistent with a white dwarf--M-dwarf binary with a circular, 4.15 d orbit. Such a wide orbit implies that there is no mass transfer in the system, as the M-dwarf would not fill its Roche lobe (which would be $> 2$~R$_{\sun}$ for this period and typical M dwarf masses). In other words, WD~J204259.71+152108.06 is not a CV.

\subsection{X-rays}
\label{xrays}

With the nature of WD~J204259.71+152108.06 yet unclear, we also analysed X-ray data for this system. An XMM-Newton observation was performed on 2019 May 20. Here we discuss only the EPIC/pn data, taken with the medium filter. We use standard products from the XMM-Newton pipeline, yielding a total on-source time of $\sim 25$ ksec.

We fit the pn data with a power-law model over the energy range 0.4--10 keV in Xspec, allowing a free absorption component from the Tuebingen absorption model (TBabs). We find that this model
provides an excellent fit to the data, with a photon index $\Gamma=2.75\pm0.06$ and 
$N_H$ of  $(1.07^{+0.12}_{-0.09}) \times 10^{21}$ cm$^{-2}$ (90\% confidence intervals for both parameters), with $\chi^2$/d.o.f. = 97/101. The unabsorbed flux from this model is $(2.8\pm0.1) \times 10^{-12}$ erg s$^{-1}$ cm$^{-2}$.

We also examined the standard products pn light curve, and find that there is no evidence for substantial variability. In detail, we used the lcstats tool to produce re-binned light curves on a variety of timescales from the initial binning of 1.46 seconds to a light curve rebinned by a factor of 256 to 373.76 seconds.  In all cases, the excess variance (i.e., the variance after subtracting the variance expected due to Poisson noise) is approximately zero. At a binning of 1.46 seconds, the 3$\sigma$ upper limit on rms fractional variation is 18\%, while at 373.76 seconds, the 3$\sigma$ upper limit on variability is 4.1\%. We examined the source's power spectrum with a range of binnings and found no statistically significant bins.


\section{Discussion}\label{sec:discussion}

\subsection{Distribution of chance alignments}
\label{sec:chances}

\begin{figure}
\includegraphics[width=\columnwidth]{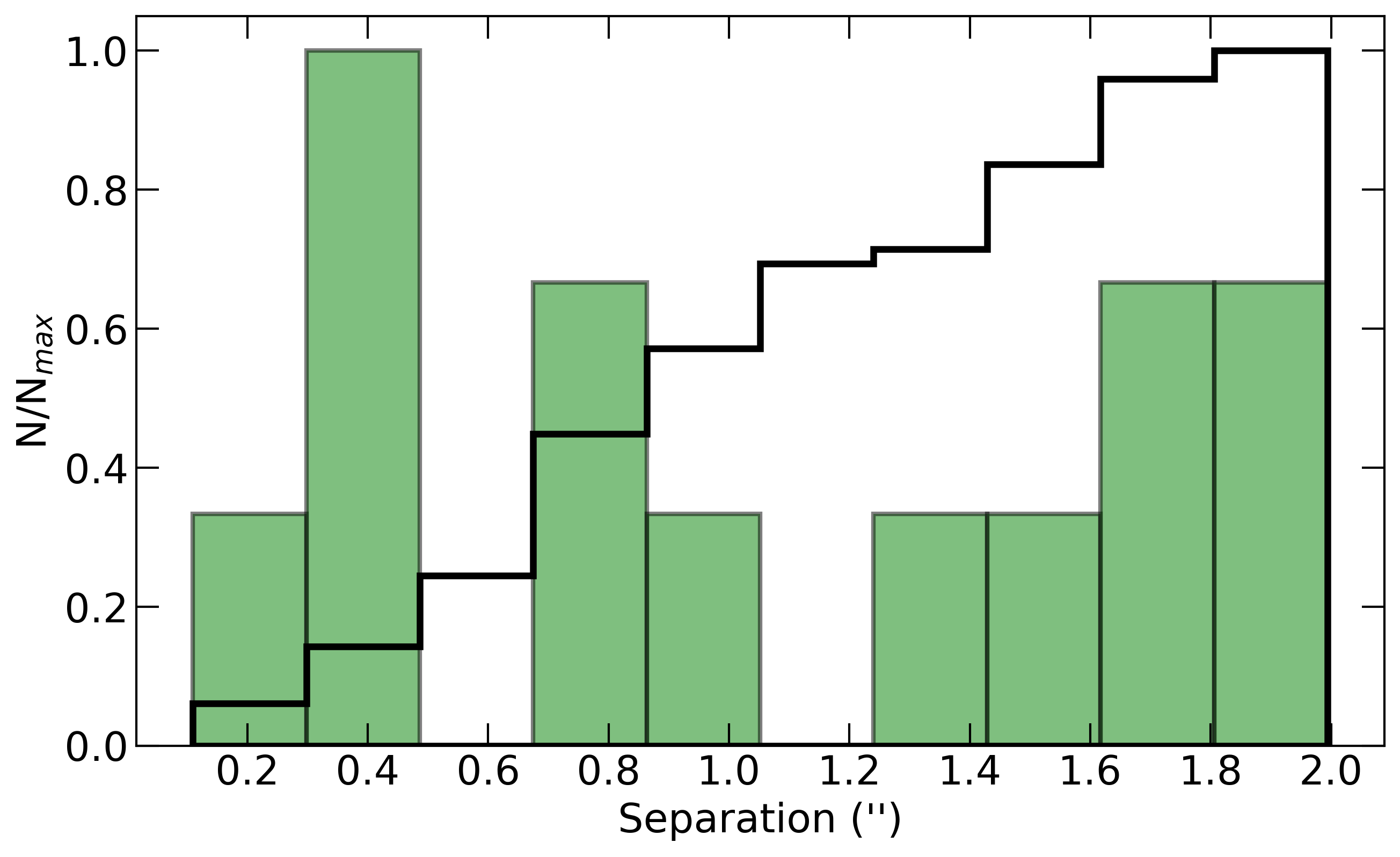}
    \caption{The green filled bars show the separation between the VLASS source and the 13 white dwarf candidates. The black line shows a histogram of the separation of all the chance alignments found when shifting the VLASS coordinates and repeating the cross-match multiple times to avoid low-number statistics (see text for details). As expected, the number of chance alignments increases with increasing separation. Both number histograms have been normalised to the maximum bin value $N_\mathrm{max}$ to facilitate direct comparison.}
    \label{fig:chancep}
\end{figure}

Even for unresolved separations, a match between optical and radio coordinates does not imply that the two sources are associated, as chance alignments are a possibility. This depends, primarily, on the sky density of sources in both catalogues. To estimate the probability of a chance alignment with a background source, we follow an approach similar to that of \citet{Lepine2007}, which uses the input catalogues themselves to characterise chance alignments, thus well describing their inherent source density. Following this approach, we repeat the match between the white dwarf candidate catalogue and VLASS, displacing the VLASS right ascension by an amount $\Delta_{\alpha}$. This change of coordinates from their real value removes any truly associated radio/optical sources, such that any resulting matches are all chance alignments. Given the low spatial density of sources in the catalogues, a single $\Delta_{\alpha}$ always results in a low ($< 25$) number of matches. Hence, to obtain a statistically significant sample of chance alignments, we used many values of $\Delta_{\alpha}$ ($10-50''$ in $10''$ steps, $75-175''$ in $25''$ steps, $200-1000''$ in $100''$ steps) and combined all resulting matches. This results in the chance alignment distribution shown in Figure~\ref{fig:chancep}. As expected, the number of chance alignments increases with separation, indicating a higher probability of chance alignment as separation increases. The separation distribution of the 13 real matches shows an excess of sources at small separation between optical and radio (Fig.~\ref{fig:chancep}) compared to the expected distribution of chance alignments, perhaps implying that some of these sources are real associations.

\subsection{The nature of WD~J204259.71+152108.06}

The only system in our search for which we could not rule out that the white dwarf and radio source are associated was WD~J204259.71+152108.06. Spectroscopic follow-up has showed this to be a white dwarf and M dwarf binary, where the white dwarf has a mass of $0.5^{+0.2}_{-0.1}$~M$_{\sun}$ and the companion is consistent with M2.5 type. Time-series spectroscopy has showed that the period is 4.15~days, much too long for a CV, from which radio emission could be expected.

Follow-up high-speed optical photometry revealed no pulses that could be associated with emission on the spin of the white dwarf, as seen for white dwarf pulsars like AR~Sco. We also did not detect flaring activity, expected from magnetic propellers like AE~Aqr. In fact, there is no indication of short-term or periodic optical variability, only long-term variability of the order of 0.2~mag in timescales of months to years. Variation on the orbital period is not expected given the relatively long period.

Ruling out a CV, a white dwarf pulsar, and a magnetic propeller excludes all possible radio emission channels that have been previously observed for white dwarfs. The possibility of a Jupiter-Io system is also very unlikely considering the predictions of \citet{WillesWu2004} for peak radio emission. Additionally, we do not find the radio emission to be strongly circularly polarised, which would be expected if the emission were due to the electron cyclotron maser mechanism.

Another possibility is that the radio emission can be attributed to the M dwarf itself. Low-frequency radio emission has been found to be ubiquitous across the M-dwarf main sequence \citep{Hallinan2008, Callingham2021}. The mechanism for coherent emission is believed to be electron cyclotron maser from the low-density regions above the star's magnetic poles \citep{Hallinan2006, Hallinan2008}. Incoherent gyrosynchrotron emission is another mechanism \citet{Gudel2002}, in particular for cooler systems. The radio flux density of this source is $\approx 1 \times 10^{-25}$~erg~s$^{-1}$~cm$^{-2}$~Hz$^{-1}$ in all measured bands. The G\"{u}del-Benz relation \citep{Guedel1993}, an empirical relationship between the radio and X-ray emission of active stars, suggests that $\log F_\mathrm{X}/F_\mathrm{R} \leq 15.5$, where $F_\mathrm{X}$ is the X-ray flux measured in erg~s$^{-1}$~cm$^{-2}$ and $F_\mathrm{R}$ is the radio flux density in erg~s$^{-1}$~cm$^{-2}$~Hz$^{-1}$. The analysis of XMM-Newton data in Section~\ref{xrays} found a 0.4--10 keV X-ray flux of $2.8 \times 10^{-12}$~erg~s$^{-1}$~cm$^{-2}$. We note the X-ray flux from an independent archival Swift/XRT measurement in the 1SXPS catalog is very similar \citep{Evans+14}, suggesting that this is an accurate measurement of the source's X-ray flux. Considering these two measurements, the radio and X-ray emissions are at first glance consistent with an M-dwarf, as $\log F_\mathrm{X}/F_\mathrm{R} \approx 13.4$. However, observations have shown that X-ray luminosity of active M-dwarfs rarely exceeds $\sim10^{-3}$ times the bolometric luminosity \citep[see e.g.][in particular the right panel of fig. 8]{Magaudda2022}. The binary has a 2MASS $K = 14.21\pm0.04$ mag \citep{2mass}, which unlike the optical or ultraviolet flux should be dominated by the M-dwarf. Assuming a $K$-band bolometric correction of 2.61 mag, typical of M2.5 dwarfs \citep[e.g.][]{Pecaut2013, Mann2015}, we estimate a bolometric flux of $\sim 5.0 \times 10^{-12}$ erg s$^{-1}$ cm$^{-2}$, giving a distance-independent $L_X/L_\mathrm{bol} \approx 0.6$. While the exact value is dependent on the X-ray band and bolometric corrections used, this is at least two orders of magnitude larger than expected for chromospheric emission from the M dwarf. Hence chromospheric emission can be ruled out as the primary source of the X-ray emission. Similarly, although the distance to the source is not well constrained (as discussed further below), the estimated range of distances would suggest a radio luminosity of $\approx 10^{18}$~erg~s$^{-1}$~Hz$^{-1}$, higher than typical for an active M-dwarf.

Concerning the distance estimate, one important factor for WD~J204259.71+152108.06 is that, even though it has a large and highly significant {\it Gaia} parallax ($\varpi = 5.64\pm0.48$ mas), the astrometric fit is rather poor (RUWE=7.3), implying that the parallax may not be accurate. A large RUWE value indicates that the photocentre of the source has wobbled during the timespan of the {\it Gaia} observations. Likely reasons are variability, multiplicity, or background sources. Variability is unlikely to play a role, as the source is not variable enough to be flagged as such by {\it Gaia}. Although WD~J204259.71+152108.06 is a binary, its separation is expected to be too small to cause a large wobble---the projected separation at 180~pc would be a mere 0.3~mas, lower than the per epoch astrometric accuracy expected for a Gaia source of its brightness \citep{Holl2023}. Conceivably, the unusual nature of the binary, with two components with extremely different intrinsic colours that cross over in brightness within the \emph{Gaia} band, together with a barely resolved semi-major axis, could lead to a poor astrometric solution, but this is speculative. If variability and multiplicity are not what causes the high RUWE, that leaves a background source as a possibility, i.e. a chance alignment.

The distribution of chance alignments derived in Section~\ref{sec:chances}, indicates that six out of 248 mock-matches show separations $\leq 0.4''$ like WD~J204259.71+152108.06, hence we estimate the chance alignment probability to be $2.1^{+0.7}_{-1.0}$ per cent (uncertainties were calculated via bootstraping). Therefore, we cannot rule out an unfortunate chance alignment between WD~J204259.71+152108.06 and a background source that would account for the radio and X-ray emission, and could potentially also contribute sufficient optical emission to affect the \emph{Gaia} astrometry. 

\begin{figure}
\centering
\includegraphics[width=\columnwidth]{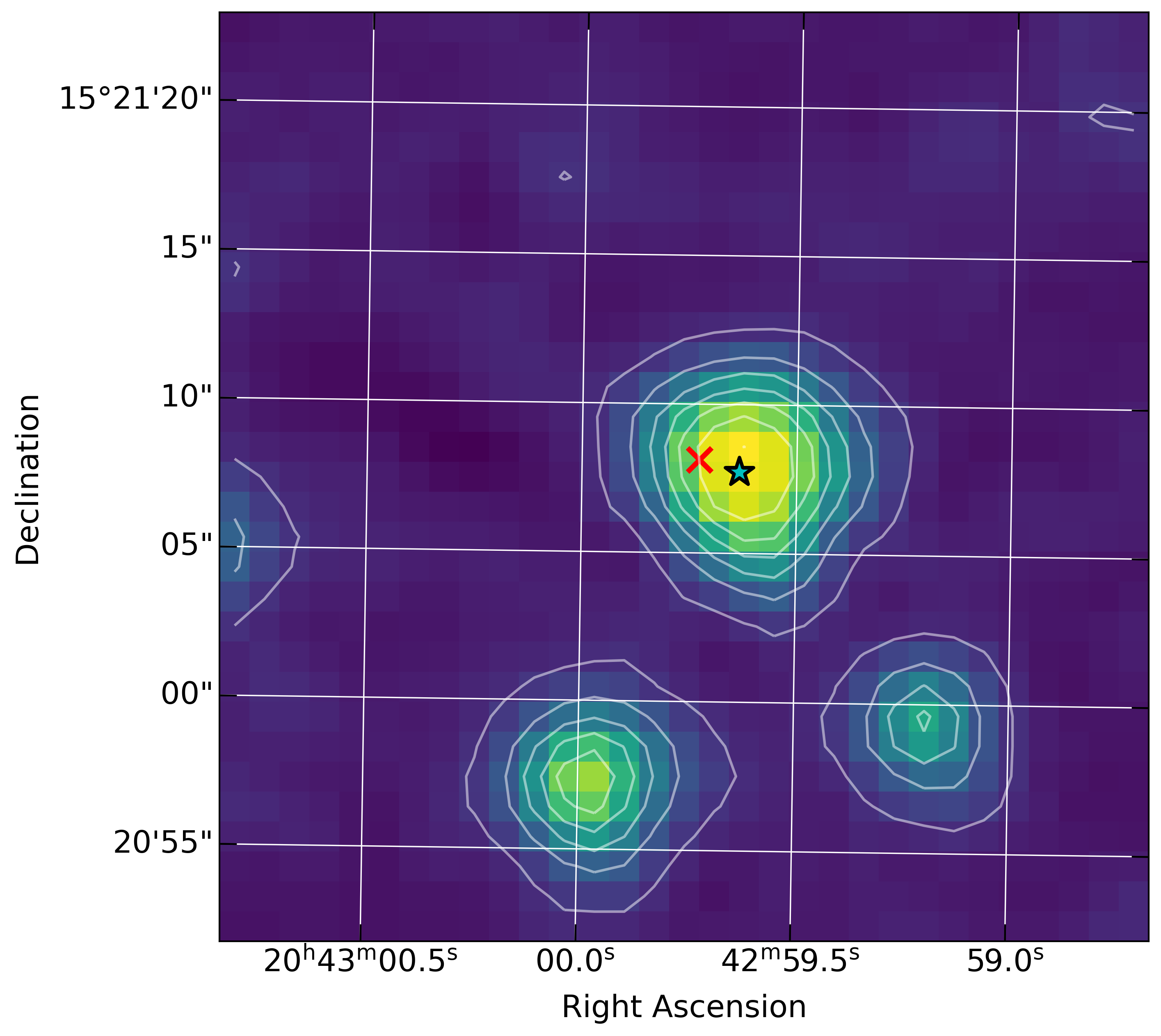}
    \caption{DSS cutout at the location of WD~J204259.71+152108.0. The cyan star shows the coordinates propagated to 1951 (the epoch of the image), whereas the red cross shows the 2016 coordinates where we would expect the quasar to be. Although there is no obvious sign of a background quasar in the image, the image scale ($1.0''$ per pixel) is comparable to the separation.}
    \label{fig:dss}
\end{figure}

One way to check the chance alignment scenario is to use the expected proper motion of the white dwarf--M dwarf binary, which should have been more separated from a putative background source in the past. We inspected images from the Digitized Sky Survey (DSS)\footnote{https://irsa.ipac.caltech.edu/data/DSS/}, which date back to the 1950s, to check whether there is indication of another source near the position of the binary. In these images, assuming the proper motions reported by {\it Gaia} are accurate and apply to the binary, there would be a $\sim 1.4''$ offset between the binary and a fixed background source. Fig.~\ref{fig:dss} shows a DSS cutout with the coordinates at the DSS and {\it Gaia} epochs indicated. There is no detection of a background source, but given the resolution of the image of $1.0''$ per pixel, comparable to the expected separation, it is not possible to rule out that the two sources are confused in the DSS image. Therefore, a chance alignment remains a possibility.

Assuming that the {\it Gaia} parallax of WD~J204259.71+152108.06 is accurate would imply a distance of $\sim 180$ pc. This distance gives $M_K \approx 8.0$ mag, corresponding to an untenably cooler star than we infer from optical spectroscopy, more like an M5V \citep[e.g.][]{Pecaut2013, Cifuentes2020}. The latter paper suggests a typical $M_K = 6.17\pm0.49$ for an M2.5V, which would imply a distance $400\pm100$ pc. Similarly for the white dwarf, we can rely on the $u$ magnitude to estimate a parallax-independent distance. SDSS reports $u = 16.65\pm0.01$, and at our derived values of $T_\mathrm{eff}$ and $\log~g$, cooling models suggest $M_u \approx 7.8$ \citep{Bedard2020}, implying a distance of also $\approx 400$~pc. Not only is this distance over a factor of two larger than the parallax distance, it is also significantly larger than the distance inferred by our spectroscopic fit ($284^{+13}_{-3}$~pc). This suggests there is additional flux contributing to the spectrum, which could be due to a contaminating background source. The residuals from our spectroscopic fit show no significant slope or features, so any additional flux would likely have no strong features (such as a BL Lacertae-type blazars). The WISE photometry for WD~J204259.71+152108.06 \citep{Cutri2021} gives $W1-W2 = 0.42\pm0.04$ and $W1-W3 = 1.76\pm0.28$, potentially redder than expected for a single M dwarf \citep{Theissen2014}, which could be consistent with the contribution of a background blazar. On the other hand, there is no detection in WISE band $W4$, which is unusual for blazars detected in the other WISE bands and with red $W1-W3$ colours (e.g., \citealt{Massaro2016}). In addition, we note that WD~J204259.71+152108.06 is located 3.7\arcmin\ from a Fermi-LAT GeV $\gamma$-ray source \citep{Abdollahi2022}. As the Fermi source has a 95\% positional semi-major axis of 2.7\arcmin, there is probably no relationship between the $\gamma$-ray source and WD~J204259.71+152108.06. However, if a future Fermi-LAT data release showed a refined position more consistent with our potential optical blend, then that would make the presence of a background blazar more likely.

In summary, a background source could explain the poor astrometric fit and the possible additional flux in the spectra. It could also be responsible for the observed long-term variability.
The inferred radio spectral index of $\alpha \approx -0.3$ is not atypical for radio galaxies \citep[e.g.][]{Zajavek2019}. Similarly, the X-ray photon index of $\Gamma  = 2.75\pm0.06$, although higher than the average value for quasars of $\Gamma  = 1.6-1.9$, is not unseen for low-redshift quasars \citep[e.g.][]{Shehata2021}.

Given how bright the source is in the radio, future very long baseline interferometric observations should be able to pin down its radio position with sufficiency accuracy to definitively rule out or confirm its association with the white dwarf--M dwarf binary.

\subsection{The paucity of bright radio emission from white dwarfs}

Out of the initial 13 matches between the {\it Gaia} white dwarf candidate catalogue and VLASS, we find that at least seven are chance alignments. Five were found to be chance alignments from analysis of deep optical images from DECam and Pan-STARRS, which revealed the radio source to be a background faint source rather than the candidate white dwarf. The remaining two chance alignments could only be revealed with high-resolution radio imaging that provided more precise radio coordinates, which were found to be inconsistent with the position of two now confirmed white dwarfs (WD~J121604.90-281909.67 and WD~J182050.14+110832.09). Three additional matches could also be chance alignments; for those, the candidates were found to be main sequence stars rather than white dwarfs (WD~J083802.17+145802.98, WD~J120358.90-001241.07, and WD~J124520.10-351755.53) and were removed from our analysis. WD~J120358.90-001241.07 has the largest separation between radio and optical coordinates in the sample, hence the association between the radio emission and the main sequence star is indeed rather unlikely. The separations are much smaller for WD~J083802.17+145802.98 ($0.5''$) and WD~J124520.10-351755.53 ($0.19''$), the latter of which actually shows the smallest separation between optical and VLASS positions in all the sample. Determining whether these two main sequence stars are indeed radio emitters is outside the scope of this work. Additionally, two out of the 13 candidate white dwarfs with a radio match were found to be quasars.

\begin{figure}
    \includegraphics[width=\columnwidth]{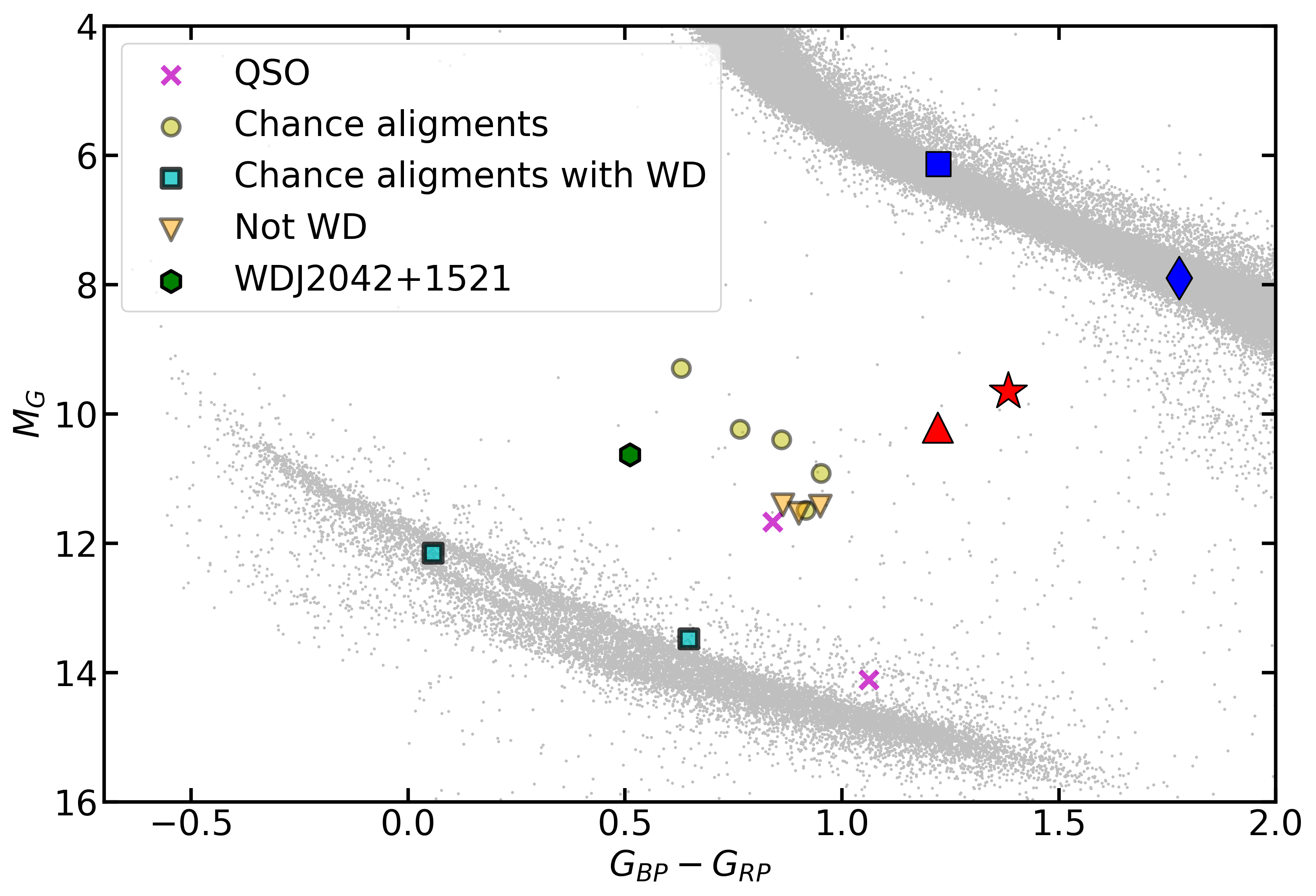}
    \caption{{\it Gaia} absolute $G$ magnitude as a function of $G_{BP} - G_{RP}$ colour for the 13 identified matches. Two were found to be quasars (magenta crosses), five are chance alignments with a candidate white dwarf (yellow circles), two chance alignments with a confirmed white dwarf (cyan squares), and three are not white dwarfs (orange triangles). Only WD~J204259.71+152108.06 (green hexagon) remains as a possible radio-emitting white dwarf. The magnetic propellers are shown as blue symbols (AE~Aqr: square, J0240: diamond) and the binary white dwarf pulsars are shown in red (AR~Sco: star, J1912: triangle).}
    \label{fig:cmd_result}
\end{figure}

\begin{table*}
	\caption{Summary of the results of our analysis, indicating whether each VLASS match was found to be a true radio source after vetting or high-resolution imaging follow-up, and the spectroscopic class of followed-up systems}
	\label{tab:conclusions}
        \centering
	\begin{tabular}{ccccccc} 
		\hline
   		\hline
		Name & Chance coincidence & Spectroscopic class & Verdict \\
		\hline
            WDJ032623.40-243623.72 & Yes & - & Candidate white dwarf is not a radio source. \\
            WDJ083802.17+145802.98 & - & FGK & Candidate is not a white dwarf. \\ 
            WDJ105211.93-335559.93 & No & QSO & Candidate is not a white dwarf. \\ 
            WDJ105223.24+313012.81 & No & QSO & Candidate is not a white dwarf. \\ 
            WDJ120358.90-001241.07 & - & FGK star & Candidate is not a white dwarf. \\ 
            WDJ121604.90-281909.67 & Yes & WD & Candidate is a white dwarf, but not a radio source, \\ 
            & & & as revealed by high-resolution radio imaging. \\
            WDJ124520.10-351755.53 & - & FGK star & Candidate is not a white dwarf. \\ 
            WDJ132423.32-255746.37 & Yes & - & Candidate white dwarf is not a radio source. \\
            WDJ182050.14+110832.09 & Yes & WD & Candidate is a white dwarf, but not a radio source \\ 
            & & & as revealed by high-resolution radio imaging. \\ 
            WDJ182112.17+204801.17 & Yes & - & Candidate white dwarf is not a radio source. \\
            WDJ183758.54-330258.93 & Yes & - & Candidate white dwarf is not a radio source. \\
            WDJ185250.55-310839.29 & Yes & - & Candidate white dwarf is not a radio source. \\
            WDJ204259.71+152108.06 & No & WD+M & Uncertain -- see text \\ 
  		\hline
        \end{tabular}
\end{table*}

This left only one white dwarf (WD~J204259.71+152108.06) from \citet{GentileFusillo+21} with radio emission detected in VLASS. These results are summarised in Fig.~\ref{fig:cmd_result} and Table~\ref{tab:conclusions}. There are 846k white dwarf candidates in the sky region covered by VLASS (dec $> 40^{\circ}$), suggesting an extremely low ($\sim 0.0001$\%) incidence of 3~GHz radio emission above 3~mJy from white dwarfs outside of interacting systems, for which the catalogue of \citet{GentileFusillo+21} is highly complete.

This is perhaps not surprising considering that the only known types of white dwarf radio emitters are interacting binaries either not targeted in the colour-magnitude selection of \citet{GentileFusillo+21}, as is the case for white dwarf pulsars and magnetic propellers (see Fig.~\ref{fig:cmd}), or that might be excluded due to their properties. The latter case refers to CVs, which can have poor astrometric solutions due to variability and hence not pass the quality criteria of \citet{GentileFusillo+21} (see their section 2.1). Additionally, depending on the contribution of the main sequence companion and of the accretion disc, they might be outside the region selected by \citet{GentileFusillo+21} \citep[see, e.g., figure 2 in][]{Abril2020}, whose main goal was to identify single white dwarfs. Out of the CVs included in \citet{GentileFusillo+21}, there are 12 known radio emitters, but all have flux densities of the order of 0.1~mJy \citep{Barrett2020}.

If we consider also the possibility of Jupiter–Io-like emission, which has never been detected from white dwarfs but has been explored in published models \citep{WillesWu2004, WillesWu2005}, the detection threshold and frequency of VLASS might also introduce a barrier. The predicted flux densities at 3~GHz are, at most, 0.01~mJy \citep[see Figure 5 in][]{WillesWu2004}, well below VLASS's threshold.

Variability may also impose a barrier to detection. For example, AR~Sco was previously observed with the VLA and showed fluxes 4--12~mJy, above the threshold for the VLASS Epoch 1 Quick Look Catalogue, and yet it is not reported as a source. We speculate that this is a result of variability, which can cause a source to be below the detection threshold when the observations happen. It is worth noting that the radio emission properties of the known binary white dwarf pulsars and magnetic propellers are so far completely dissimilar. Whereas AE~Aqr, detected by VLASS, shows flux higher than 5~mJy, a positive spectral index, and no polarisation, the other propeller, J0240+1952, shows much lower ($< 1$~mJy) flux, a negative spectral index, and high circular polarisation \citep{Barrett2022}. AR~Sco shows strong variability in radio in many timescales, including broad pulses, whereas J1912-4410 shows no emission other than extremely narrow pulses \citep{Pelisoli+23}. Therefore, making any predictions about the expected detectability of such systems is not straightforward and further radio searches are encouraged. VLASS itself might be a useful tool for this. Our estimates of the flux density for the detected sources in epochs 1 and 2 (Table~\ref{tab:wd_vlass}) indicates that some of the detected sources (such as WD~J204259.71+152108.06) show variability, which suggests that variable sources might be detectable with VLASS under favourable circumstances. Future source catalogues based on epochs 2 and 3 might therefore reveal other variable sources that escaped from detection in epoch 1.

\section{Summary and Conclusions} \label{sec:conclusion}

We performed a cross-match between the white dwarf candidate catalogue of \citet{GentileFusillo+21} and VLASS epoch 1 \citep{Gordon+20,Gordon+21}, identifying 13 sources that matched within $2''$. Two of the white dwarf candidates turned out to be quasars according to spectroscopy, and three were found to be main sequence stars. Of the remaining eight sources, five were found to be chance alignments through analysis of deeper optical images, and two were found to be chance alignments with a white dwarf using higher resolution radio imaging. The only source for which the association between the {\it Gaia} optical source and the radio stood up to scrutiny was WD~J204259.71+152108.06.

This source was found to be a binary system with a white dwarf and an M-dwarf on a 4.1-day orbit, too wide for significant mass-transfer that could explain radio emission like that seen for CVs. We found no short-term variability that could associate this source with radio-emitting white dwarf pulsars or magnetic propellers. The lack of strong circular polarisation additionally rules out electron cyclotron maser from a Jupiter-Io analogue. The radio and X-ray properties of the source are consistent with a quasar, and we do find evidence for additional flux in our spectra that could be due to a background source. However, we find the probability of chance alignment to be low ($\sim 2$\%) given the small ($<0.4''$) separation between the radio and optical coordinates. Additionally, archival DSS images where the position of the binary would be shifted by $1.4''$ from a static background source reveal no evidence of a resolved source, but the low resolution of the DSS images ($1.0''$) makes decisive conclusions difficult.

WD~J204259.71+152108.06 could therefore be explained by a near-perfect chance alignment with a background radio source, although the lack of detection of this source in archival images as well as the low probability of such an alignment remain a puzzle. Future long-baseline radio observations can settle the matter definitively. With at most one (and quite possibly zero) genuine radio-emitting white dwarfs found in VLASS, our main conclusion is that radio emission from white dwarfs outside of interacting binaries, at least at 3~GHz and with fluxes $\gtrsim 1-3$~mJy, is extremely rare to non-existent.

\section*{Acknowledgements}

This work was conceived of and initiated at the KITP program ``White Dwarfs as Probes of the Evolution of Planets, Stars, the Milky Way and the Expanding Universe", which was supported in part by the National Science Foundation under Grant No. NSF PHY-1748958. We are grateful to the organisers of this program for spurring new collaborations.

This research was partially supported by the Munich Institute for Astro-, Particle and BioPhysics (MIAPbP) which is funded by the Deutsche Forschungsgemeinschaft (DFG, German Research Foundation) under Germany´s Excellence Strategy – EXC-2094 – 390783311. IP thanks the organisers and participants of the MIAPbP workshop Stellar Magnetic Fields from Protostars to Supernovae for helpful discussions, in particular Rob Kavanagh and Antonino Lanza.

IP acknowledges support from a Warwick Astrophysics prize post-doctoral fellowship, made possible thanks to a generous philanthropic donation, and from a Royal Society University Research Fellowship (URF\textbackslash R1\textbackslash 231496).

LC is grateful for support from the National Science Foundation (under grants AST-1751874,  AST-2107070, and AST-2205628) and NASA (under grant 80NSSC23K0497).

JS acknowledges support from the Packard Foundation.

This project has received funding from the European Research Council (ERC) under the European Union’s Horizon 2020 research and innovation programme (Grant agreement No. 101020057).

TJM thanks the late Tom Marsh for useful discussions both about J204259.71+152108.06, and many other topics over the years.

KCD acknowledges support for this work
provided by NASA through the NASA Hubble Fellowship grant
HST-HF2-51528 awarded by the Space Telescope Science Institute, which is operated by the Association of Universities for Research in Astronomy, Inc., for NASA, under contract NAS5–26555. 

VSD and ULTRACAM are funded by the Science and Technology Facilities Council (grant ST/V000853/1).

The National Radio Astronomy Observatory is a facility of the National Science Foundation operated under cooperative agreement by Associated Universities, Inc. Based on observations obtained at the Southern Astrophysical Research (SOAR) telescope, which is a joint project of the Minist\'{e}rio da Ci\^{e}ncia, Tecnologia e Inova\c{c}\~{o}es (MCTI/LNA) do Brasil, the US National Science Foundation’s NOIRLab, the University of North Carolina at Chapel Hill (UNC), and Michigan State University (MSU).
This research has made use of the International Variable Star Index (VSX) database, operated at AAVSO, Cambridge, Massachusetts, USA.  We acknowledge with thanks the variable star observations from the AAVSO International Database contributed by observers worldwide and used in this research. This work has made use of data from the European Space Agency (ESA) mission {\it Gaia} (\url{https://www.cosmos.esa.int/gaia}), processed by the {\it Gaia} Data Processing and Analysis Consortium (DPAC,\url{https://www.cosmos.esa.int/web/gaia/dpac/consortium}). Funding for the DPAC has been provided by national institutions, in particular the institutions participating in the {\it Gaia} Multilateral Agreement. This research has made use of the CIRADA cutout service at \url{http://cutouts.cirada.ca/}, operated by the Canadian Initiative for Radio Astronomy Data Analysis (CIRADA). CIRADA is funded by a grant from the Canada Foundation for Innovation 2017 Innovation Fund (Project 35999), as well as by the Provinces of Ontario, British Columbia, Alberta, Manitoba and Quebec, in collaboration with the National Research Council of Canada, the US National Radio Astronomy Observatory and Australia’s Commonwealth Scientific and Industrial Research Organisation. This research uses services and data provided by the Astro Data Lab at NSF's National Optical-Infrared Astronomy Research Laboratory. NOIRLab is operated by the Association of Universities for Research in Astronomy (AURA), Inc. under a cooperative agreement with the National Science Foundation. 

For the purpose of open access, the author has applied a Creative Commons Attribution (CC-BY) licence to any Author Accepted Manuscript version arising from this submission.


\section*{Data Availability}

All data analysed in this work can be made available upon reasonable request to the authors.



\bibliographystyle{mnras}
\bibliography{wd} 

\begin{thebibliography}{}
\makeatletter
\relax
\def\mn@urlcharsother{\let\do\@makeother \do\$\do\&\do\#\do\^\do\_\do\%\do\~}
\def\mn@doi{\begingroup\mn@urlcharsother \@ifnextchar [ {\mn@doi@} {\mn@doi@[]}}
\def\mn@doi@[#1]#2{\def\@tempa{#1}\ifx\@tempa\@empty \href {http://dx.doi.org/#2} {doi:#2}\else \href {http://dx.doi.org/#2} {#1}\fi \endgroup}
\def\mn@eprint#1#2{\mn@eprint@#1:#2::\@nil}
\def\mn@eprint@arXiv#1{\href {http://arxiv.org/abs/#1} {{\tt arXiv:#1}}}
\def\mn@eprint@dblp#1{\href {http://dblp.uni-trier.de/rec/bibtex/#1.xml} {dblp:#1}}
\def\mn@eprint@#1:#2:#3:#4\@nil{\def\@tempa {#1}\def\@tempb {#2}\def\@tempc {#3}\ifx \@tempc \@empty \let \@tempc \@tempb \let \@tempb \@tempa \fi \ifx \@tempb \@empty \def\@tempb {arXiv}\fi \@ifundefined {mn@eprint@\@tempb}{\@tempb:\@tempc}{\expandafter \expandafter \csname mn@eprint@\@tempb\endcsname \expandafter{\@tempc}}}

\bibitem[\protect\citeauthoryear{{Abdollahi} et~al.,}{{Abdollahi} et~al.}{2022}]{Abdollahi2022}
{Abdollahi} S.,  et~al., 2022, \mn@doi [The Astrophysical Journal Supplement Series] {10.3847/1538-4365/ac6751}, \href {https://ui.adsabs.harvard.edu/abs/2022ApJS..260...53A} {260, 53}

\bibitem[\protect\citeauthoryear{{Abril}, {Schmidtobreick}, {Ederoclite}  \& {L{\'o}pez-Sanjuan}}{{Abril} et~al.}{2020}]{Abril2020}
{Abril} J.,  {Schmidtobreick} L.,  {Ederoclite} A.,   {L{\'o}pez-Sanjuan} C.,  2020, \mn@doi [\mnras] {10.1093/mnrasl/slz181}, \href {https://ui.adsabs.harvard.edu/abs/2020MNRAS.492L..40A} {492, L40}

\bibitem[\protect\citeauthoryear{{Ahumada} et~al.,}{{Ahumada} et~al.}{2020}]{Ahumada2020}
{Ahumada} R.,  et~al., 2020, \mn@doi [\apjs] {10.3847/1538-4365/ab929e}, \href {https://ui.adsabs.harvard.edu/abs/2020ApJS..249....3A} {249, 3}

\bibitem[\protect\citeauthoryear{{Allard}, {Homeier}  \& {Freytag}}{{Allard} et~al.}{2013}]{Allard2013}
{Allard} F.,  {Homeier} D.,   {Freytag} B.,  2013, \memsai, \href {https://ui.adsabs.harvard.edu/abs/2013MmSAI..84.1053A} {84, 1053}

\bibitem[\protect\citeauthoryear{{Bailer-Jones}, {Rybizki}, {Fouesneau}, {Demleitner}  \& {Andrae}}{{Bailer-Jones} et~al.}{2021}]{Bailer-Jones+2021}
{Bailer-Jones} C.~A.~L.,  {Rybizki} J.,  {Fouesneau} M.,  {Demleitner} M.,   {Andrae} R.,  2021, \mn@doi [\aj] {10.3847/1538-3881/abd806}, \href {https://ui.adsabs.harvard.edu/abs/2021AJ....161..147B} {161, 147}

\bibitem[\protect\citeauthoryear{{Barrett}}{{Barrett}}{2022}]{Barrett2022}
{Barrett} P.~E.,  2022, \mn@doi [\aj] {10.3847/1538-3881/ac3ed9}, \href {https://ui.adsabs.harvard.edu/abs/2022AJ....163...58B} {163, 58}

\bibitem[\protect\citeauthoryear{{Barrett}, {Dieck}, {Beasley}, {Mason}  \& {Singh}}{{Barrett} et~al.}{2020}]{Barrett2020}
{Barrett} P.,  {Dieck} C.,  {Beasley} A.~J.,  {Mason} P.~A.,   {Singh} K.~P.,  2020, \mn@doi [Advances in Space Research] {10.1016/j.asr.2020.04.007}, \href {https://ui.adsabs.harvard.edu/abs/2020AdSpR..66.1226B} {66, 1226}

\bibitem[\protect\citeauthoryear{{Bastian}, {Dulk}  \& {Chanmugam}}{{Bastian} et~al.}{1988}]{Bastian1988}
{Bastian} T.~S.,  {Dulk} G.~A.,   {Chanmugam} G.,  1988, \mn@doi [\apj] {10.1086/165906}, \href {https://ui.adsabs.harvard.edu/abs/1988ApJ...324..431B} {324, 431}

\bibitem[\protect\citeauthoryear{{B{\'e}dard}, {Bergeron}, {Brassard}  \& {Fontaine}}{{B{\'e}dard} et~al.}{2020}]{Bedard2020}
{B{\'e}dard} A.,  {Bergeron} P.,  {Brassard} P.,   {Fontaine} G.,  2020, \mn@doi [\apj] {10.3847/1538-4357/abafbe}, \href {https://ui.adsabs.harvard.edu/abs/2020ApJ...901...93B} {901, 93}

\bibitem[\protect\citeauthoryear{{Bellm} et~al.,}{{Bellm} et~al.}{2019}]{ztf}
{Bellm} E.~C.,  et~al., 2019, \mn@doi [\pasp] {10.1088/1538-3873/aaecbe}, \href {https://ui.adsabs.harvard.edu/abs/2019PASP..131a8002B} {131, 018002}

\bibitem[\protect\citeauthoryear{{Bookbinder} \& {Lamb}}{{Bookbinder} \& {Lamb}}{1987}]{Bookbinder1987}
{Bookbinder} J.~A.,  {Lamb} D.~Q.,  1987, \mn@doi [\apjl] {10.1086/185072}, \href {https://ui.adsabs.harvard.edu/abs/1987ApJ...323L.131B} {323, L131}

\bibitem[\protect\citeauthoryear{{Callingham} et~al.,}{{Callingham} et~al.}{2021}]{Callingham2021}
{Callingham} J.~R.,  et~al., 2021, \mn@doi [Nature Astronomy] {10.1038/s41550-021-01483-0}, \href {https://ui.adsabs.harvard.edu/abs/2021NatAs...5.1233C} {5, 1233}

\bibitem[\protect\citeauthoryear{{Capitanio}, {Lallement}, {Vergely}, {Elyajouri}  \& {Monreal-Ibero}}{{Capitanio} et~al.}{2017}]{Capitanio2017}
{Capitanio} L.,  {Lallement} R.,  {Vergely} J.~L.,  {Elyajouri} M.,   {Monreal-Ibero} A.,  2017, \mn@doi [\aap] {10.1051/0004-6361/201730831}, \href {https://ui.adsabs.harvard.edu/abs/2017A&A...606A..65C} {606, A65}

\bibitem[\protect\citeauthoryear{{Chambers} et~al.,}{{Chambers} et~al.}{2016}]{ps1}
{Chambers} K.~C.,  et~al., 2016, \mn@doi [arXiv e-prints] {10.48550/arXiv.1612.05560}, \href {https://ui.adsabs.harvard.edu/abs/2016arXiv161205560C} {p. arXiv:1612.05560}

\bibitem[\protect\citeauthoryear{{Cifuentes} et~al.,}{{Cifuentes} et~al.}{2020}]{Cifuentes2020}
{Cifuentes} C.,  et~al., 2020, \mn@doi [\aap] {10.1051/0004-6361/202038295}, \href {https://ui.adsabs.harvard.edu/abs/2020A&A...642A.115C} {642, A115}

\bibitem[\protect\citeauthoryear{{Clemens}, {Crain}  \& {Anderson}}{{Clemens} et~al.}{2004}]{goodman}
{Clemens} J.~C.,  {Crain} J.~A.,   {Anderson} R.,  2004, in {Moorwood} A. F.~M.,  {Iye} M.,  eds,  Society of Photo-Optical Instrumentation Engineers (SPIE) Conference Series Vol. 5492, Ground-based Instrumentation for Astronomy. pp 331--340, \mn@doi{10.1117/12.550069}

\bibitem[\protect\citeauthoryear{{Coppejans}, {K{\"o}rding}, {Miller-Jones}, {Rupen}, {Knigge}, {Sivakoff}  \& {Groot}}{{Coppejans} et~al.}{2015}]{Coppejans2015}
{Coppejans} D.~L.,  {K{\"o}rding} E.~G.,  {Miller-Jones} J. C.~A.,  {Rupen} M.~P.,  {Knigge} C.,  {Sivakoff} G.~R.,   {Groot} P.~J.,  2015, \mn@doi [\mnras] {10.1093/mnras/stv1225}, \href {https://ui.adsabs.harvard.edu/abs/2015MNRAS.451.3801C} {451, 3801}

\bibitem[\protect\citeauthoryear{{Coppejans} et~al.,}{{Coppejans} et~al.}{2016}]{Coppejans2016}
{Coppejans} D.~L.,  et~al., 2016, \mn@doi [\mnras] {10.1093/mnras/stw2133}, \href {https://ui.adsabs.harvard.edu/abs/2016MNRAS.463.2229C} {463, 2229}

\bibitem[\protect\citeauthoryear{{Cutri} et~al.,}{{Cutri} et~al.}{2021}]{Cutri2021}
{Cutri} R.~M.,  et~al., 2021, VizieR Online Data Catalog, \href {https://ui.adsabs.harvard.edu/abs/2014yCat.2328....0C} {p. II/328}

\bibitem[\protect\citeauthoryear{{Dhillon} et~al.,}{{Dhillon} et~al.}{2007}]{ultracam}
{Dhillon} V.~S.,  et~al., 2007, \mn@doi [\mnras] {10.1111/j.1365-2966.2007.11881.x}, \href {https://ui.adsabs.harvard.edu/abs/2007MNRAS.378..825D} {378, 825}

\bibitem[\protect\citeauthoryear{{Dhillon} et~al.,}{{Dhillon} et~al.}{2014}]{ultraspec}
{Dhillon} V.~S.,  et~al., 2014, \mn@doi [\mnras] {10.1093/mnras/stu1660}, \href {https://ui.adsabs.harvard.edu/abs/2014MNRAS.444.4009D} {444, 4009}

\bibitem[\protect\citeauthoryear{{Drake} et~al.,}{{Drake} et~al.}{2009}]{catalina}
{Drake} A.~J.,  et~al., 2009, \mn@doi [\apj] {10.1088/0004-637X/696/1/870}, \href {https://ui.adsabs.harvard.edu/abs/2009ApJ...696..870D} {696, 870}

\bibitem[\protect\citeauthoryear{{Drlica-Wagner} et~al.,}{{Drlica-Wagner} et~al.}{2021a}]{delve}
{Drlica-Wagner} A.,  et~al., 2021a, \mn@doi [\apjs] {10.3847/1538-4365/ac079d}, \href {https://ui.adsabs.harvard.edu/abs/2021ApJS..256....2D} {256, 2}

\bibitem[\protect\citeauthoryear{{Drlica-Wagner} et~al.,}{{Drlica-Wagner} et~al.}{2021b}]{Drlica-Wagner+21}
{Drlica-Wagner} A.,  et~al., 2021b, \mn@doi [\apjs] {10.3847/1538-4365/ac079d}, \href {https://ui.adsabs.harvard.edu/abs/2021ApJS..256....2D} {256, 2}

\bibitem[\protect\citeauthoryear{{Dubus}, {Otulakowska-Hypka}  \& {Lasota}}{{Dubus} et~al.}{2018}]{Dubus+2018}
{Dubus} G.,  {Otulakowska-Hypka} M.,   {Lasota} J.-P.,  2018, \mn@doi [\aap] {10.1051/0004-6361/201833372}, \href {https://ui.adsabs.harvard.edu/abs/2018A&A...617A..26D} {617, A26}

\bibitem[\protect\citeauthoryear{{Evans} et~al.,}{{Evans} et~al.}{2014}]{Evans+14}
{Evans} P.~A.,  et~al., 2014, \mn@doi [\apjs] {10.1088/0067-0049/210/1/8}, \href {https://ui.adsabs.harvard.edu/abs/2014ApJS..210....8E} {210, 8}

\bibitem[\protect\citeauthoryear{{Gaia Collaboration} et~al.,}{{Gaia Collaboration} et~al.}{2023}]{GaiaDR3}
{Gaia Collaboration} et~al., 2023, \mn@doi [\aap] {10.1051/0004-6361/202243940}, \href {https://ui.adsabs.harvard.edu/abs/2023A&A...674A...1G} {674, A1}

\bibitem[\protect\citeauthoryear{{Gentile Fusillo} et~al.,}{{Gentile Fusillo} et~al.}{2021}]{GentileFusillo+21}
{Gentile Fusillo} N.~P.,  et~al., 2021, \mn@doi [\mnras] {10.1093/mnras/stab2672}, \href {https://ui.adsabs.harvard.edu/abs/2021MNRAS.508.3877G} {508, 3877}

\bibitem[\protect\citeauthoryear{{Gordon} et~al.,}{{Gordon} et~al.}{2020}]{Gordon+20}
{Gordon} Y.~A.,  et~al., 2020, \mn@doi [Research Notes of the American Astronomical Society] {10.3847/2515-5172/abbe23}, \href {https://ui.adsabs.harvard.edu/abs/2020RNAAS...4..175G} {4, 175}

\bibitem[\protect\citeauthoryear{{Gordon} et~al.,}{{Gordon} et~al.}{2021}]{Gordon+21}
{Gordon} Y.~A.,  et~al., 2021, \mn@doi [\apjs] {10.3847/1538-4365/ac05c0}, \href {https://ui.adsabs.harvard.edu/abs/2021ApJS..255...30G} {255, 30}

\bibitem[\protect\citeauthoryear{{Greisen}}{{Greisen}}{2003}]{Greisen03}
{Greisen} E.~W.,  2003, in {Heck} A.,  ed.,  Astrophysics and Space Science Library Vol. 285, Information Handling in Astronomy - Historical Vistas. p.~109, \mn@doi{10.1007/0-306-48080-8_7}

\bibitem[\protect\citeauthoryear{{G{\"u}del}}{{G{\"u}del}}{2002}]{Gudel2002}
{G{\"u}del} M.,  2002, \mn@doi [\araa] {10.1146/annurev.astro.40.060401.093806}, \href {https://ui.adsabs.harvard.edu/abs/2002ARA&A..40..217G} {40, 217}

\bibitem[\protect\citeauthoryear{{Guedel} \& {Benz}}{{Guedel} \& {Benz}}{1993}]{Guedel1993}
{Guedel} M.,  {Benz} A.~O.,  1993, \mn@doi [\apjl] {10.1086/186766}, \href {https://ui.adsabs.harvard.edu/abs/1993ApJ...405L..63G} {405, L63}

\bibitem[\protect\citeauthoryear{{Hallinan}, {Antonova}, {Doyle}, {Bourke}, {Brisken}  \& {Golden}}{{Hallinan} et~al.}{2006}]{Hallinan2006}
{Hallinan} G.,  {Antonova} A.,  {Doyle} J.~G.,  {Bourke} S.,  {Brisken} W.~F.,   {Golden} A.,  2006, \mn@doi [\apj] {10.1086/508678}, \href {https://ui.adsabs.harvard.edu/abs/2006ApJ...653..690H} {653, 690}

\bibitem[\protect\citeauthoryear{{Hallinan}, {Antonova}, {Doyle}, {Bourke}, {Lane}  \& {Golden}}{{Hallinan} et~al.}{2008}]{Hallinan2008}
{Hallinan} G.,  {Antonova} A.,  {Doyle} J.~G.,  {Bourke} S.,  {Lane} C.,   {Golden} A.,  2008, \mn@doi [\apj] {10.1086/590360}, \href {https://ui.adsabs.harvard.edu/abs/2008ApJ...684..644H} {684, 644}

\bibitem[\protect\citeauthoryear{{Harayama}, {Terada}, {Ishida}, {Hayashi}, {Bamba}  \& {Tashiro}}{{Harayama} et~al.}{2013}]{Harayama+2013}
{Harayama} A.,  {Terada} Y.,  {Ishida} M.,  {Hayashi} T.,  {Bamba} A.,   {Tashiro} M.~S.,  2013, \mn@doi [\pasj] {10.1093/pasj/65.4.73}, \href {https://ui.adsabs.harvard.edu/abs/2013PASJ...65...73H} {65, 73}

\bibitem[\protect\citeauthoryear{{Hermes} et~al.,}{{Hermes} et~al.}{2017}]{Hermes2017}
{Hermes} J.~J.,  et~al., 2017, \mn@doi [\apjs] {10.3847/1538-4365/aa8bb5}, \href {https://ui.adsabs.harvard.edu/abs/2017ApJS..232...23H} {232, 23}

\bibitem[\protect\citeauthoryear{{Hewitt} et~al.,}{{Hewitt} et~al.}{2020}]{Hewitt2020}
{Hewitt} D.~M.,  et~al., 2020, \mn@doi [\mnras] {10.1093/mnras/staa1747}, \href {https://ui.adsabs.harvard.edu/abs/2020MNRAS.496.2542H} {496, 2542}

\bibitem[\protect\citeauthoryear{{Holl} et~al.,}{{Holl} et~al.}{2023}]{Holl2023}
{Holl} B.,  et~al., 2023, \mn@doi [Astronomy and Astrophysics] {10.1051/0004-6361/202244161}, \href {https://ui.adsabs.harvard.edu/abs/2023A&A...674A..10H} {674, A10}

\bibitem[\protect\citeauthoryear{{Katz}}{{Katz}}{2017}]{Katz2017}
{Katz} J.~I.,  2017, \mn@doi [\apj] {10.3847/1538-4357/835/2/150}, \href {https://ui.adsabs.harvard.edu/abs/2017ApJ...835..150K} {835, 150}

\bibitem[\protect\citeauthoryear{{Koester}}{{Koester}}{2010}]{Koester2010}
{Koester} D.,  2010, \memsai, \href {https://ui.adsabs.harvard.edu/abs/2010MmSAI..81..921K} {81, 921}

\bibitem[\protect\citeauthoryear{{K{\"o}rding}, {Rupen}, {Knigge}, {Fender}, {Dhawan}, {Templeton}  \& {Muxlow}}{{K{\"o}rding} et~al.}{2008}]{Kording2008}
{K{\"o}rding} E.,  {Rupen} M.,  {Knigge} C.,  {Fender} R.,  {Dhawan} V.,  {Templeton} M.,   {Muxlow} T.,  2008, \mn@doi [Science] {10.1126/science.1155492}, \href {https://ui.adsabs.harvard.edu/abs/2008Sci...320.1318K} {320, 1318}

\bibitem[\protect\citeauthoryear{{Kroupa}}{{Kroupa}}{2001}]{Kroupa2001}
{Kroupa} P.,  2001, \mn@doi [\mnras] {10.1046/j.1365-8711.2001.04022.x}, \href {https://ui.adsabs.harvard.edu/abs/2001MNRAS.322..231K} {322, 231}

\bibitem[\protect\citeauthoryear{{Lacy} et~al.,}{{Lacy} et~al.}{2019}]{Lacy+19}
{Lacy} M.,  et~al., 2019, {VLASS Project Memo \#13: Pilot and Quick Look Data Release (v2)}

\bibitem[\protect\citeauthoryear{{Lauffer}, {Romero}  \& {Kepler}}{{Lauffer} et~al.}{2018}]{Lauffer+2018}
{Lauffer} G.~R.,  {Romero} A.~D.,   {Kepler} S.~O.,  2018, \mn@doi [\mnras] {10.1093/mnras/sty1925}, \href {https://ui.adsabs.harvard.edu/abs/2018MNRAS.480.1547L} {480, 1547}

\bibitem[\protect\citeauthoryear{{L{\'e}pine} \& {Bongiorno}}{{L{\'e}pine} \& {Bongiorno}}{2007}]{Lepine2007}
{L{\'e}pine} S.,  {Bongiorno} B.,  2007, \mn@doi [\aj] {10.1086/510333}, \href {https://ui.adsabs.harvard.edu/abs/2007AJ....133..889L} {133, 889}

\bibitem[\protect\citeauthoryear{{Li}, {Ferrario}  \& {Wickramasinghe}}{{Li} et~al.}{1998}]{Li1998}
{Li} J.,  {Ferrario} L.,   {Wickramasinghe} D.,  1998, \mn@doi [\apjl] {10.1086/311546}, \href {https://ui.adsabs.harvard.edu/abs/1998ApJ...503L.151L} {503, L151}

\bibitem[\protect\citeauthoryear{{Lindegren} et~al.,}{{Lindegren} et~al.}{2021}]{Lindegren2021}
{Lindegren} L.,  et~al., 2021, \mn@doi [\aap] {10.1051/0004-6361/202039653}, \href {https://ui.adsabs.harvard.edu/abs/2021A&A...649A...4L} {649, A4}

\bibitem[\protect\citeauthoryear{{Lomb}}{{Lomb}}{1976}]{Lomb1976}
{Lomb} N.~R.,  1976, \mn@doi [\apss] {10.1007/BF00648343}, \href {https://ui.adsabs.harvard.edu/abs/1976Ap&SS..39..447L} {39, 447}

\bibitem[\protect\citeauthoryear{{Luo}, {Zhao}, {Zhao}  \& {et al.}}{{Luo} et~al.}{2022}]{Luo2022}
{Luo} A.~L.,  {Zhao} Y.~H.,  {Zhao} G.,   {et al.} 2022, VizieR Online Data Catalog, \href {https://ui.adsabs.harvard.edu/abs/2022yCat.5156....0L} {p. V/156}

\bibitem[\protect\citeauthoryear{{Magaudda}, {Stelzer}, {Raetz}, {Klutsch}, {Salvato}  \& {Wolf}}{{Magaudda} et~al.}{2022}]{Magaudda2022}
{Magaudda} E.,  {Stelzer} B.,  {Raetz} S.,  {Klutsch} A.,  {Salvato} M.,   {Wolf} J.,  2022, \mn@doi [\aap] {10.1051/0004-6361/202141617}, \href {https://ui.adsabs.harvard.edu/abs/2022A&A...661A..29M} {661, A29}

\bibitem[\protect\citeauthoryear{{Magnier} et~al.,}{{Magnier} et~al.}{2020}]{Magnier+20}
{Magnier} E.~A.,  et~al., 2020, \mn@doi [\apjs] {10.3847/1538-4365/abb82a}, \href {https://ui.adsabs.harvard.edu/abs/2020ApJS..251....6M} {251, 6}

\bibitem[\protect\citeauthoryear{{Mann}, {Feiden}, {Gaidos}, {Boyajian}  \& {von Braun}}{{Mann} et~al.}{2015}]{Mann2015}
{Mann} A.~W.,  {Feiden} G.~A.,  {Gaidos} E.,  {Boyajian} T.,   {von Braun} K.,  2015, \mn@doi [\apj] {10.1088/0004-637X/804/1/64}, \href {https://ui.adsabs.harvard.edu/abs/2015ApJ...804...64M} {804, 64}

\bibitem[\protect\citeauthoryear{{Marsh} et~al.,}{{Marsh} et~al.}{2016}]{Marsh2016}
{Marsh} T.~R.,  et~al., 2016, \mn@doi [\nat] {10.1038/nature18620}, \href {https://ui.adsabs.harvard.edu/abs/2016Natur.537..374M} {537, 374}

\bibitem[\protect\citeauthoryear{{Massaro} \& {D'Abrusco}}{{Massaro} \& {D'Abrusco}}{2016}]{Massaro2016}
{Massaro} F.,  {D'Abrusco} R.,  2016, \mn@doi [The Astrophysical Journal] {10.3847/0004-637X/827/1/67}, \href {https://ui.adsabs.harvard.edu/abs/2016ApJ...827...67M} {827, 67}

\bibitem[\protect\citeauthoryear{{Meintjes} \& {Venter}}{{Meintjes} \& {Venter}}{2005}]{Meintjes2005}
{Meintjes} P.~J.,  {Venter} L.~A.,  2005, \mn@doi [\mnras] {10.1111/j.1365-2966.2005.09045.x}, \href {https://ui.adsabs.harvard.edu/abs/2005MNRAS.360..573M} {360, 573}

\bibitem[\protect\citeauthoryear{{Mohan} \& {Rafferty}}{{Mohan} \& {Rafferty}}{2015}]{PyBDSF}
{Mohan} N.,  {Rafferty} D.,  2015, {PyBDSF: Python Blob Detection and Source Finder}, Astrophysics Source Code Library, record ascl:1502.007

\bibitem[\protect\citeauthoryear{{Nelemans}, {Yungelson}, {Portegies Zwart}  \& {Verbunt}}{{Nelemans} et~al.}{2001}]{Nelemans+2001}
{Nelemans} G.,  {Yungelson} L.~R.,  {Portegies Zwart} S.~F.,   {Verbunt} F.,  2001, \mn@doi [\aap] {10.1051/0004-6361:20000147}, \href {https://ui.adsabs.harvard.edu/abs/2001A&A...365..491N} {365, 491}

\bibitem[\protect\citeauthoryear{{O'Brien} et~al.,}{{O'Brien} et~al.}{2024}]{OBrien2024}
{O'Brien} M.~W.,  et~al., 2024, \mn@doi [\mnras] {10.1093/mnras/stad3773}, \href {https://ui.adsabs.harvard.edu/abs/2024MNRAS.527.8687O} {527, 8687}

\bibitem[\protect\citeauthoryear{{Padoan} \& {Nordlund}}{{Padoan} \& {Nordlund}}{2002}]{Padoan+2002}
{Padoan} P.,  {Nordlund} {\r{A}}.,  2002, \mn@doi [\apj] {10.1086/341790}, \href {https://ui.adsabs.harvard.edu/abs/2002ApJ...576..870P} {576, 870}

\bibitem[\protect\citeauthoryear{{Patterson}}{{Patterson}}{1979}]{Patterson1979}
{Patterson} J.,  1979, \mn@doi [\apj] {10.1086/157582}, \href {https://ui.adsabs.harvard.edu/abs/1979ApJ...234..978P} {234, 978}

\bibitem[\protect\citeauthoryear{{Pecaut} \& {Mamajek}}{{Pecaut} \& {Mamajek}}{2013}]{Pecaut2013}
{Pecaut} M.~J.,  {Mamajek} E.~E.,  2013, \mn@doi [\apjs] {10.1088/0067-0049/208/1/9}, \href {https://ui.adsabs.harvard.edu/abs/2013ApJS..208....9P} {208, 9}

\bibitem[\protect\citeauthoryear{{Pelisoli} et~al.,}{{Pelisoli} et~al.}{2022}]{Pelisoli+2022a}
{Pelisoli} I.,  et~al., 2022, \mn@doi [\mnras] {10.1093/mnrasl/slab116}, \href {https://ui.adsabs.harvard.edu/abs/2022MNRAS.509L..31P} {509, L31}

\bibitem[\protect\citeauthoryear{{Pelisoli} et~al.,}{{Pelisoli} et~al.}{2023}]{Pelisoli+23}
{Pelisoli} I.,  et~al., 2023, \mn@doi [Nature Astronomy] {10.1038/s41550-023-01995-x}, \href {https://ui.adsabs.harvard.edu/abs/2023NatAs.tmp..120P} {}

\bibitem[\protect\citeauthoryear{{Perley}, {Chandler}, {Butler}  \& {Wrobel}}{{Perley} et~al.}{2011}]{Perley2011}
{Perley} R.~A.,  {Chandler} C.~J.,  {Butler} B.~J.,   {Wrobel} J.~M.,  2011, \mn@doi [\apjl] {10.1088/2041-8205/739/1/L1}, \href {https://ui.adsabs.harvard.edu/abs/2011ApJ...739L...1P} {739, L1}

\bibitem[\protect\citeauthoryear{{Potter} \& {Buckley}}{{Potter} \& {Buckley}}{2018}]{PotterBuckley2018}
{Potter} S.~B.,  {Buckley} D. A.~H.,  2018, \mn@doi [\mnras] {10.1093/mnras/sty2407}, \href {https://ui.adsabs.harvard.edu/abs/2018MNRAS.481.2384P} {481, 2384}

\bibitem[\protect\citeauthoryear{{Pretorius} et~al.,}{{Pretorius} et~al.}{2021}]{Pretorius+2021}
{Pretorius} M.~L.,  et~al., 2021, \mn@doi [\mnras] {10.1093/mnras/stab498}, \href {https://ui.adsabs.harvard.edu/abs/2021MNRAS.503.3692P} {503, 3692}

\bibitem[\protect\citeauthoryear{{Price-Whelan}, {Hogg}, {Foreman-Mackey}  \& {Rix}}{{Price-Whelan} et~al.}{2017}]{Price-Whelan2017}
{Price-Whelan} A.~M.,  {Hogg} D.~W.,  {Foreman-Mackey} D.,   {Rix} H.-W.,  2017, \mn@doi [The Astrophysical Journal] {10.3847/1538-4357/aa5e50}, \href {https://ui.adsabs.harvard.edu/abs/2017ApJ...837...20P} {837, 20}

\bibitem[\protect\citeauthoryear{{Prochaska} et~al.,}{{Prochaska} et~al.}{2020}]{Prochaska2020}
{Prochaska} J.,  et~al., 2020, \mn@doi [The Journal of Open Source Software] {10.21105/joss.02308}, \href {https://ui.adsabs.harvard.edu/abs/2020JOSS....5.2308P} {5, 2308}

\bibitem[\protect\citeauthoryear{{Ridder}, {Heinke}, {Sivakoff}  \& {Hughes}}{{Ridder} et~al.}{2023}]{Ridder2023}
{Ridder} M.~E.,  {Heinke} C.~O.,  {Sivakoff} G.~R.,   {Hughes} A.~K.,  2023, \mn@doi [\mnras] {10.1093/mnras/stad038}, \href {https://ui.adsabs.harvard.edu/abs/2023MNRAS.519.5922R} {519, 5922}

\bibitem[\protect\citeauthoryear{{Ritter} \& {Kolb}}{{Ritter} \& {Kolb}}{2003}]{Ritter2003}
{Ritter} H.,  {Kolb} U.,  2003, \mn@doi [\aap] {10.1051/0004-6361:20030330}, \href {https://ui.adsabs.harvard.edu/abs/2003A&A...404..301R} {404, 301}

\bibitem[\protect\citeauthoryear{{Russell} et~al.,}{{Russell} et~al.}{2016}]{Russell2016}
{Russell} T.~D.,  et~al., 2016, \mn@doi [\mnras] {10.1093/mnras/stw1238}, \href {https://ui.adsabs.harvard.edu/abs/2016MNRAS.460.3720R} {460, 3720}

\bibitem[\protect\citeauthoryear{{Scargle}}{{Scargle}}{1982}]{Scargle1982}
{Scargle} J.~D.,  1982, \mn@doi [\apj] {10.1086/160554}, \href {https://ui.adsabs.harvard.edu/abs/1982ApJ...263..835S} {263, 835}

\bibitem[\protect\citeauthoryear{{Scaringi}}{{Scaringi}}{2014}]{Scaringi2014}
{Scaringi} S.,  2014, \mn@doi [\mnras] {10.1093/mnras/stt2270}, \href {https://ui.adsabs.harvard.edu/abs/2014MNRAS.438.1233S} {438, 1233}

\bibitem[\protect\citeauthoryear{{Schreiber}, {Belloni}, {G{\"a}nsicke}, {Parsons}  \& {Zorotovic}}{{Schreiber} et~al.}{2021}]{Schreiber2021}
{Schreiber} M.~R.,  {Belloni} D.,  {G{\"a}nsicke} B.~T.,  {Parsons} S.~G.,   {Zorotovic} M.,  2021, \mn@doi [Nature Astronomy] {10.1038/s41550-021-01346-8}, \href {https://ui.adsabs.harvard.edu/abs/2021NatAs...5..648S} {5, 648}

\bibitem[\protect\citeauthoryear{{Schwope}, {Marsh}, {Standke}, {Pelisoli}, {Potter}, {Buckley}, {Munday}  \& {Dhillon}}{{Schwope} et~al.}{2023}]{Schwope2023}
{Schwope} A.,  {Marsh} T.~R.,  {Standke} A.,  {Pelisoli} I.,  {Potter} S.,  {Buckley} D.,  {Munday} J.,   {Dhillon} V.,  2023, \mn@doi [\aap] {10.1051/0004-6361/202346589}, \href {https://ui.adsabs.harvard.edu/abs/2023A&A...674L...9S} {674, L9}

\bibitem[\protect\citeauthoryear{{Shehata}, {Misra}, {Osman}, {Shalabiea}  \& {Hayman}}{{Shehata} et~al.}{2021}]{Shehata2021}
{Shehata} S.~M.,  {Misra} R.,  {Osman} A.~M.~I.,  {Shalabiea} O.~M.,   {Hayman} Z.~M.,  2021, \mn@doi [Journal of High Energy Astrophysics] {10.1016/j.jheap.2021.04.003}, \href {https://ui.adsabs.harvard.edu/abs/2021JHEAp..31...37S} {31, 37}

\bibitem[\protect\citeauthoryear{{Skrutskie} et~al.,}{{Skrutskie} et~al.}{2006}]{2mass}
{Skrutskie} M.~F.,  et~al., 2006, \mn@doi [\aj] {10.1086/498708}, \href {https://ui.adsabs.harvard.edu/abs/2006AJ....131.1163S} {131, 1163}

\bibitem[\protect\citeauthoryear{{Stanway}, {Marsh}, {Chote}, {G{\"a}nsicke}, {Steeghs}  \& {Wheatley}}{{Stanway} et~al.}{2018}]{Stanway+18}
{Stanway} E.~R.,  {Marsh} T.~R.,  {Chote} P.,  {G{\"a}nsicke} B.~T.,  {Steeghs} D.,   {Wheatley} P.~J.,  2018, \mn@doi [\aap] {10.1051/0004-6361/201732380}, \href {https://ui.adsabs.harvard.edu/abs/2018A&A...611A..66S} {611, A66}

\bibitem[\protect\citeauthoryear{{Takata}, {Yang}  \& {Cheng}}{{Takata} et~al.}{2017}]{Takata2017}
{Takata} J.,  {Yang} H.,   {Cheng} K.~S.,  2017, \mn@doi [\apj] {10.3847/1538-4357/aa9b33}, \href {https://ui.adsabs.harvard.edu/abs/2017ApJ...851..143T} {851, 143}

\bibitem[\protect\citeauthoryear{{Theissen} \& {West}}{{Theissen} \& {West}}{2014}]{Theissen2014}
{Theissen} C.~A.,  {West} A.~A.,  2014, \mn@doi [The Astrophysical Journal] {10.1088/0004-637X/794/2/146}, \href {https://ui.adsabs.harvard.edu/abs/2014ApJ...794..146T} {794, 146}

\bibitem[\protect\citeauthoryear{{Thorstensen}}{{Thorstensen}}{2020}]{Thorstensen2020}
{Thorstensen} J.~R.,  2020, \mn@doi [\aj] {10.3847/1538-3881/aba7c7}, \href {https://ui.adsabs.harvard.edu/abs/2020AJ....160..151T} {160, 151}

\bibitem[\protect\citeauthoryear{{Tody}}{{Tody}}{1986}]{Tody1986}
{Tody} D.,  1986, in {Crawford} D.~L.,  ed.,  Society of Photo-Optical Instrumentation Engineers (SPIE) Conference Series Vol. 627, Instrumentation in astronomy VI. p.~733, \mn@doi{10.1117/12.968154}

\bibitem[\protect\citeauthoryear{{Tonry} et~al.,}{{Tonry} et~al.}{2018}]{atlas}
{Tonry} J.~L.,  et~al., 2018, \mn@doi [\pasp] {10.1088/1538-3873/aabadf}, \href {https://ui.adsabs.harvard.edu/abs/2018PASP..130f4505T} {130, 064505}

\bibitem[\protect\citeauthoryear{{Toonen}, {Hollands}, {G{\"a}nsicke}  \& {Boekholt}}{{Toonen} et~al.}{2017}]{Toonen+2017}
{Toonen} S.,  {Hollands} M.,  {G{\"a}nsicke} B.~T.,   {Boekholt} T.,  2017, \mn@doi [\aap] {10.1051/0004-6361/201629978}, \href {https://ui.adsabs.harvard.edu/abs/2017A&A...602A..16T} {602, A16}

\bibitem[\protect\citeauthoryear{{Warner}}{{Warner}}{1995}]{Warner1995}
{Warner} B.,  1995, {Cataclysmic variable stars}.
~ Vol. 28

\bibitem[\protect\citeauthoryear{{Willes} \& {Wu}}{{Willes} \& {Wu}}{2004}]{WillesWu2004}
{Willes} A.~J.,  {Wu} K.,  2004, \mn@doi [\mnras] {10.1111/j.1365-2966.2004.07363.x}, \href {https://ui.adsabs.harvard.edu/abs/2004MNRAS.348..285W} {348, 285}

\bibitem[\protect\citeauthoryear{{Willes} \& {Wu}}{{Willes} \& {Wu}}{2005}]{WillesWu2005}
{Willes} A.~J.,  {Wu} K.,  2005, \mn@doi [\aap] {10.1051/0004-6361:20040417}, \href {https://ui.adsabs.harvard.edu/abs/2005A&A...432.1091W} {432, 1091}

\bibitem[\protect\citeauthoryear{{Woosley} \& {Heger}}{{Woosley} \& {Heger}}{2015}]{WoosleyHeger2015}
{Woosley} S.~E.,  {Heger} A.,  2015, \mn@doi [\apj] {10.1088/0004-637X/810/1/34}, \href {https://ui.adsabs.harvard.edu/abs/2015ApJ...810...34W} {810, 34}

\bibitem[\protect\citeauthoryear{{Zaja{\v{c}}ek} et~al.,}{{Zaja{\v{c}}ek} et~al.}{2019}]{Zajavek2019}
{Zaja{\v{c}}ek} M.,  et~al., 2019, \mn@doi [\aap] {10.1051/0004-6361/201833388}, \href {https://ui.adsabs.harvard.edu/abs/2019A&A...630A..83Z} {630, A83}

\bibitem[\protect\citeauthoryear{{du Plessis}, {Venter}, {Wadiasingh}, {Harding}, {Buckley}, {Potter}  \& {Meintjes}}{{du Plessis} et~al.}{2022}]{Plessis2022}
{du Plessis} L.,  {Venter} C.,  {Wadiasingh} Z.,  {Harding} A.~K.,  {Buckley} D. A.~H.,  {Potter} S.~B.,   {Meintjes} P.~J.,  2022, \mn@doi [\mnras] {10.1093/mnras/stab3595}, \href {https://ui.adsabs.harvard.edu/abs/2022MNRAS.510.2998D} {510, 2998}

\makeatother
\end{thebibliography}


\clearpage

\appendix

\section{Radio detected CVs}

\begin{table*}
	\caption{Cataclysmic variables with radio detections reported in the literature. We give the Simbad name, the class following the classification scheme of \citet{Ritter2003} (AM = polar, IP = intermediate polar, Na = fast nova, DN = dwarf nova, NL = nova-like, DS = detached system, here a pre-CV), the radio band and the highest flux reported in the literature, and the references. }
	\label{tab:radiodets}
        \centering
	\begin{tabular}{cccccc}
		\hline
        Name & Type & Radio band & Flux [$\mu$Jy] & Reference \\
        \hline
          SS Cyg & DN & X & 1100 & \citet{Kording2008, Russell2016}\\
          RW Sex & NL & C & 82$\pm$23 & \citet{Coppejans2015, Hewitt2020}\\
          TT Ari & NL & C & 239.1 & \citet{Coppejans2015}\\
          V603 Aql & Na & C & 233$\pm$36 & \citet{Coppejans2015, Hewitt2020}\\
          Z Cam & DN & X & 40.3$\pm$5.2 & \citet{Coppejans2016}\\
          RX And & DN & X & 19.6$\pm$4.4 & \citet{Coppejans2016}\\
          SU UMa & DN & X & 58.1$\pm$5.7 & \citet{Coppejans2016}\\
          YZ Cnc & DN & X & 26.8$\pm$5.2 & \citet{Coppejans2016}\\
          U Gem & DN & X & 12.7$\pm$2.8 & \citet{Coppejans2016}\\
          EQ Cet & AM & K & 96$\pm$30 & \citet{Barrett2020}\\
          Cas 1 & IP & C & 21$\pm$10 & \citet{Barrett2020}\\
          FL Cet & AM & X & 11$\pm$5 & \citet{Barrett2020}\\
          BS Tri & AM & X & 57$\pm$9 & \citet{Barrett2020}\\
          EF Eri & AM & X & 87$\pm$15 & \citet{Barrett2020}\\
          UZ For & AM & C & 78$\pm$9 & \citet{Barrett2020}\\
          Tau 4 & AM? & X & 105$\pm$32 & \citet{Barrett2020}\\
          LW Cam & AM & X & 50$\pm$4 & \citet{Barrett2020}\\
          VV Pup & AM & X & 79$\pm$14 & \citet{Barrett2020}\\
          FR Lyn & AM & X & 28$\pm$4 & \citet{Barrett2020}\\
          RXJ0859.1+0537 (Hya 1) & AM & X & 6$\pm$6 & \citet{Barrett2020}\\
          HS0922+1333 & AM & X & 8$\pm$5 & \citet{Barrett2020}\\
          WX LMi & AM & C & 73$\pm$12 & \citet{Barrett2020}\\
          ST LMi & AM & X & 153$\pm$12 & \citet{Barrett2020}\\
          AR UMa & AM & C & 489$\pm$16 & \citet{Barrett2020}\\
          EU UMa & AM & X & 39$\pm$5 & \citet{Barrett2020}\\
          V1043 Cen & AM & X & 20$\pm$5 & \citet{Barrett2020}\\
          CRTS J150354.0-220710 (J1503-2207) & AM & X & 29$\pm$5 & \citet{Barrett2020}\\
          BM CrB & AM & X & 43$\pm$15 & \citet{Barrett2020}\\
          MR Ser & AM & C & 239$\pm$17 & \citet{Barrett2020}\\
          MQ Dra & AM & X & 17$\pm$4 & \citet{Barrett2020}\\
          AP CrB & AM & X & 24$\pm$4 & \citet{Barrett2020}\\
          SDSS J162608.16+332827.7 (Her 1) & AM & K & 48$\pm$17 & \citet{Barrett2020}\\
          V1007 Her & AM & X & 38$\pm$9 & \citet{Barrett2020}\\
          V1323 Her & IP & C & 43$\pm$9 & \citet{Barrett2020}\\
          AM Her & AM & K & 476$\pm$83 & \citet{Barrett2020}\\
          V603 Aql & Na & X & 32$\pm$7 & \citet{Barrett2020}\\
          V1432 Aql & AM & X & 15$\pm$5 & \citet{Barrett2020}\\
          2MASS J19551247+0045365 (J1955+0045) & AM & X & 79$\pm$5 & \citet{Barrett2020}\\
          QQ Vul & AM & K & 92$\pm$39 & \citet{Barrett2020}\\
          HU Aqr & AM & X & 44$\pm$13 & \citet{Barrett2020}\\
          V388 Peg & AM & X & 34$\pm$5 & \citet{Barrett2020}\\
          IM Eri & NL & L & 99$\pm$26 & \citet{Hewitt2020}\\
          V3885 Sgr & NL & L & 256$\pm$25 & \citet{Hewitt2020}\\
          V2400 Oph & IP & S & 640$\pm$120 & \citet{Ridder2023}\\
          QS Vir & DS & S & 970$\pm$160 & \citet{Ridder2023}\\
          \hline
    \end{tabular}
\end{table*}

\section{Periodograms of long-term photometric data}

\begin{figure}
    \includegraphics[width=\columnwidth]{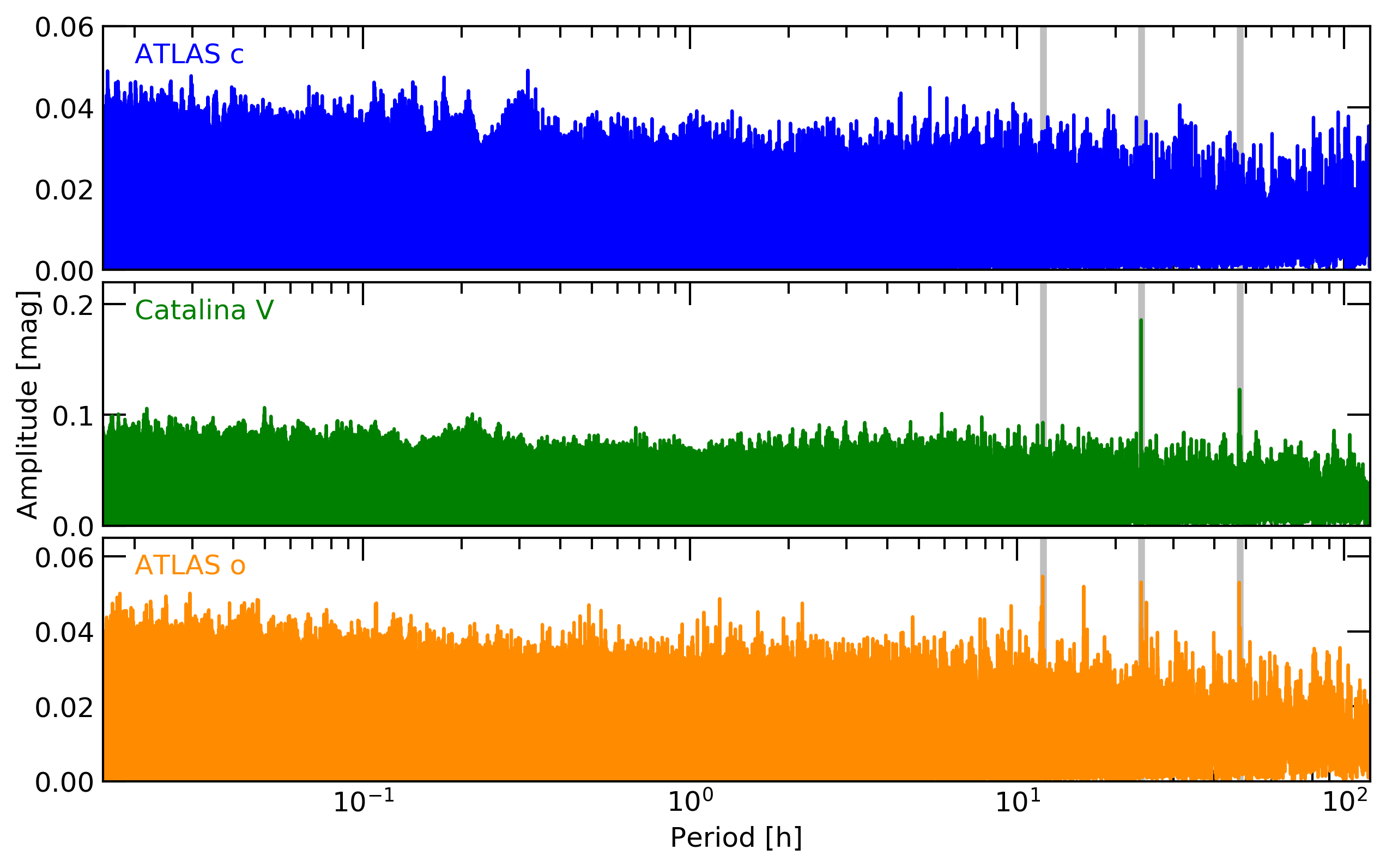}
    \caption{Periodograms for WD~J121604.90-281909.67 for ATLAS $c$ (top), Catalina $V$ (middle) and ATLAS $o$ (bottom). The grey vertical lines indicate the location of some one-day aliases that are visible. No genuine periods are identified.}
    \label{fig:j1216ft}
\end{figure}

\begin{figure}
    \includegraphics[width=\columnwidth]{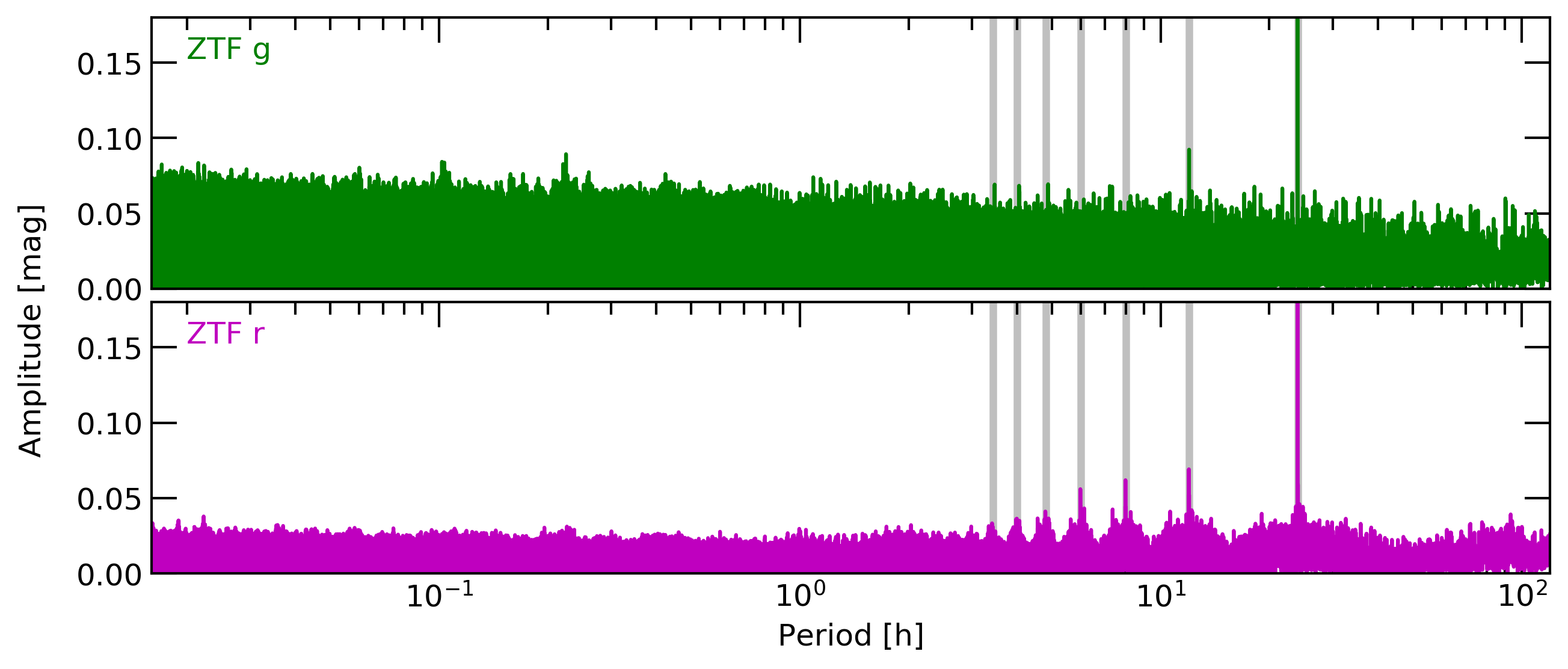}
    \caption{Periodograms for the ZTF light curves of WD~J182050.14+110832.09. The grey vertical lines indicate one-day aliases, which are the only visible peaks.}
    \label{fig:j1820ft}
\end{figure}

\begin{figure}
    \includegraphics[width=\columnwidth]{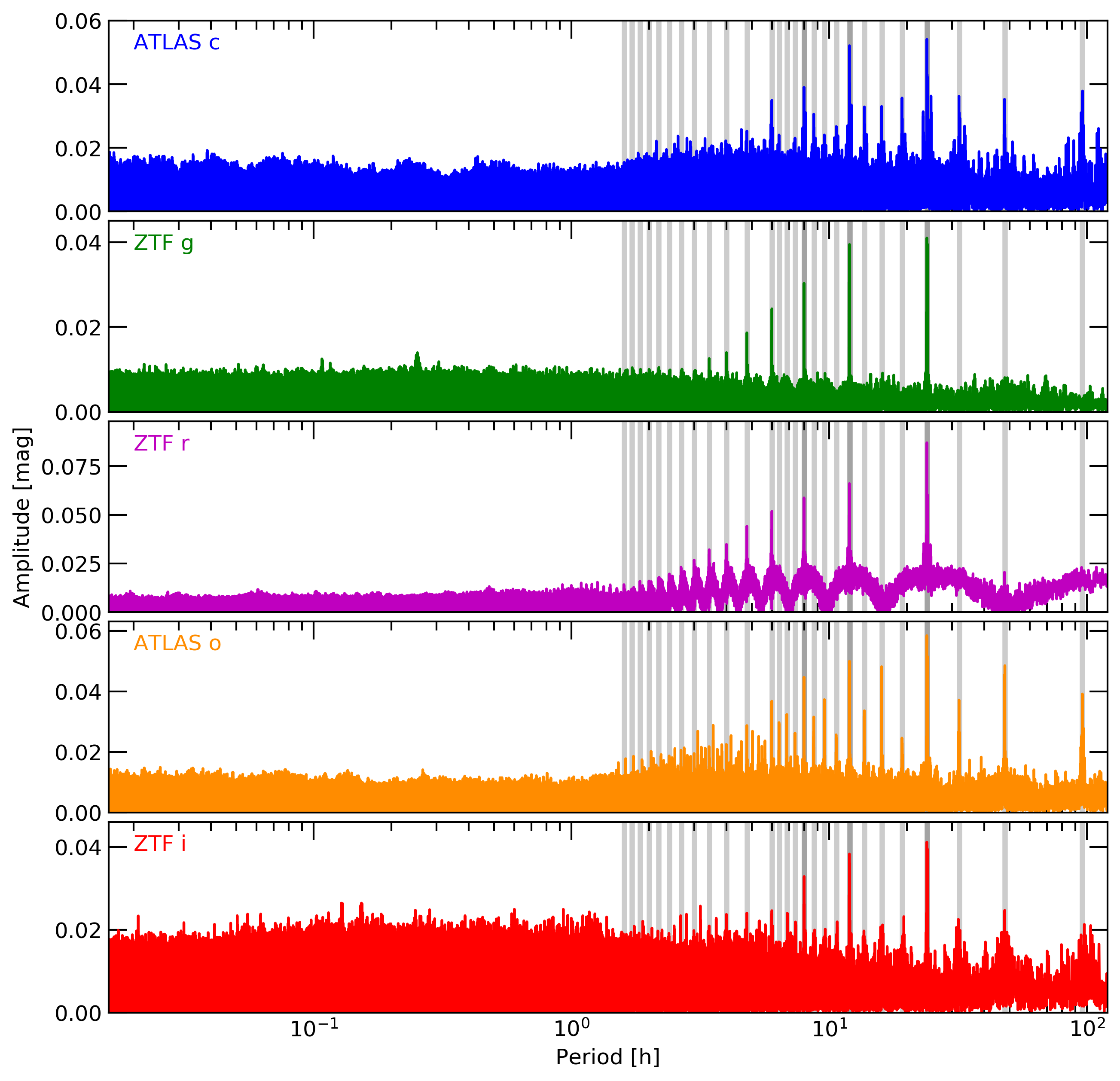}
    \caption{Periodograms for WD~J204259.71+152108.06. A number of one-day aliases are present and highlighted as the grey lines, which mark 96 h/$i$ and 24 h/$i$ for $j=[1,15]$}
    \label{fig:j2042ft}
\end{figure}

\section{Follow-up photometric data}

\begin{figure*}
    \includegraphics[width=\textwidth]{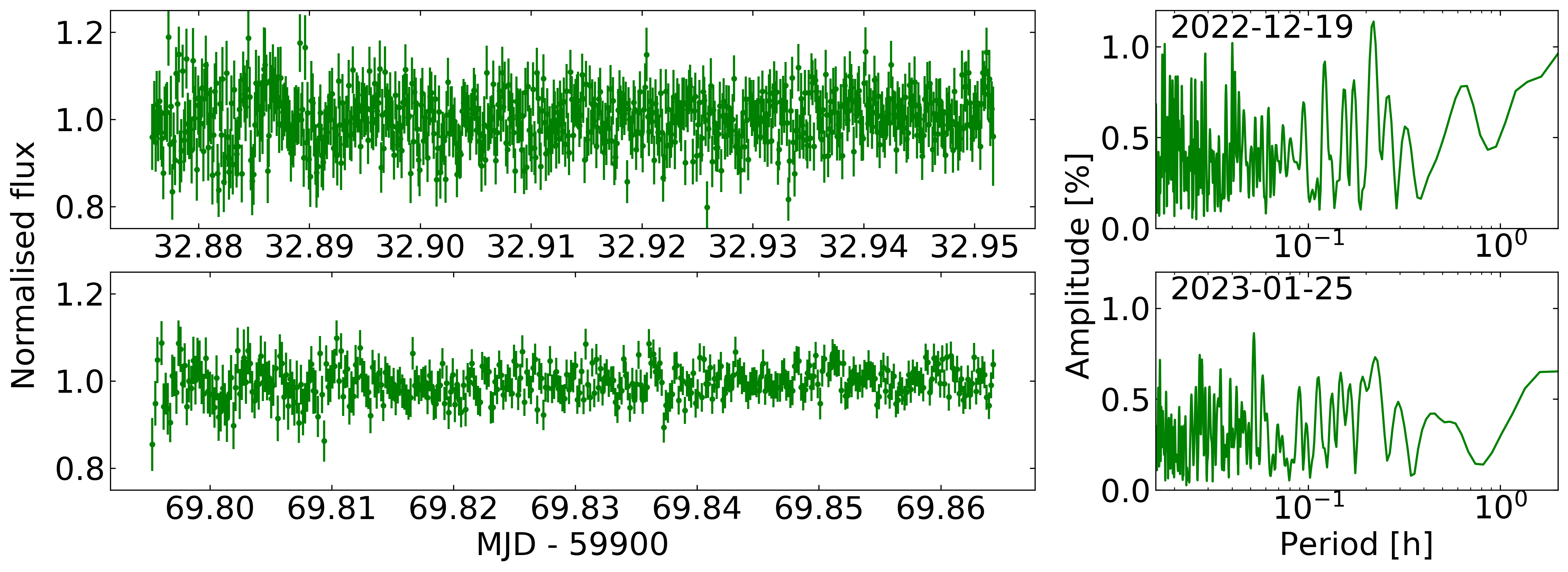}
    \caption{Light curves (left panels) and periodograms (right panels) for the $g_s$ ULTRASPEC data obtained for WD~J121604.90-281909.67 on 2022 December 19 (top) and 2023 January 25 (bottom). We find no evidence for periodic or aperiodic variability.}
    \label{fig:j1216uspec}
\end{figure*}

\begin{figure*}
    \includegraphics[width=\textwidth]{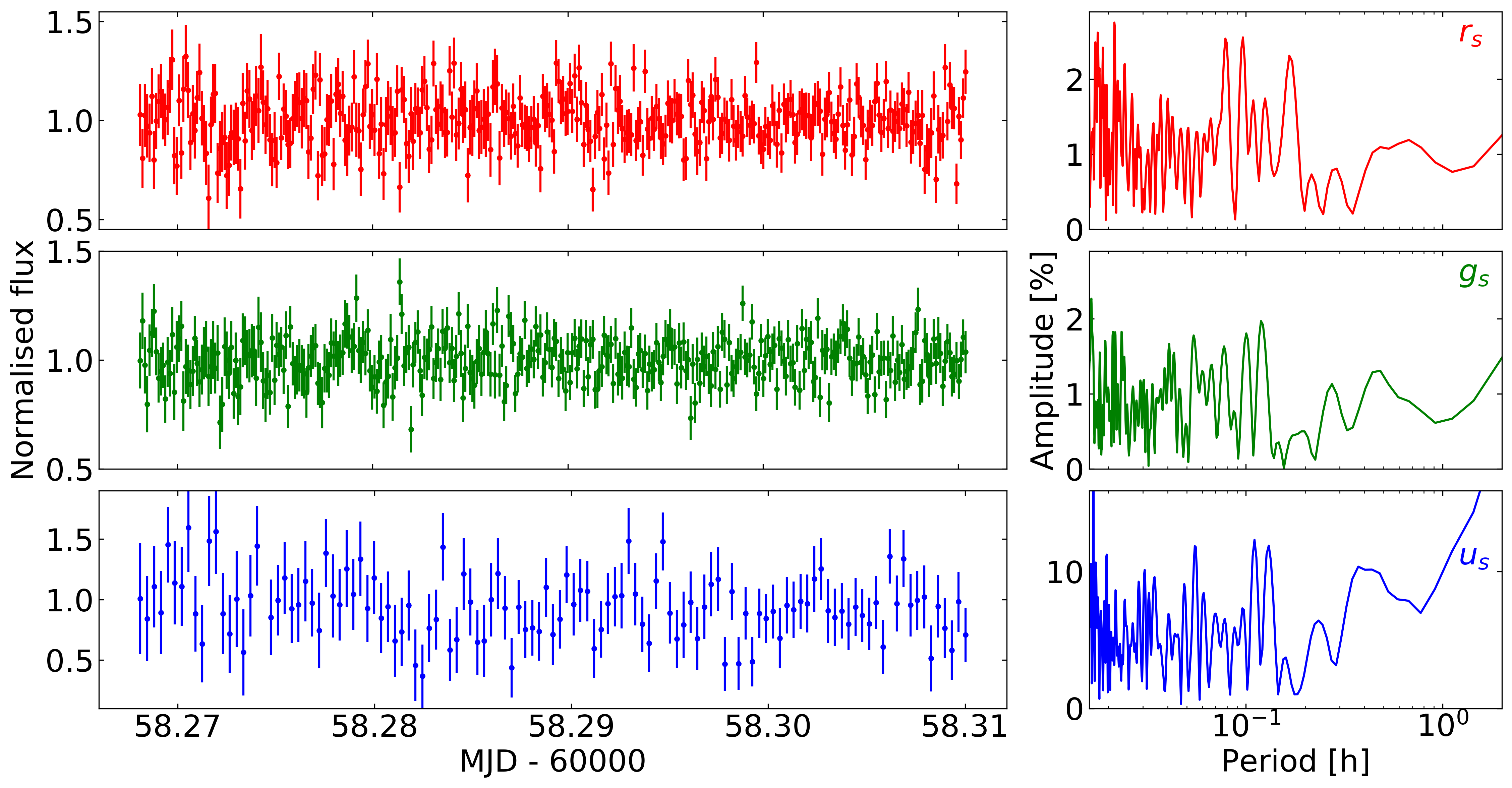}
    \caption{Light curves (left panels) and periodograms (right panels) for the $r_s$, $g_s$, and $u_s$ ULTRACAM data obtained for WD~J182050.14+110832.09. The periodogram shows no significant peaks, with the only strong peak in $u_s$ appearing at 1~minute, i.e. two times the cadence. The light curves also show no evidence for stochastic variability, with the scatter being consistent with Gaussian noise.}
    \label{fig:j1820ucam}
\end{figure*}

\begin{figure*}
    \includegraphics[width=\textwidth]{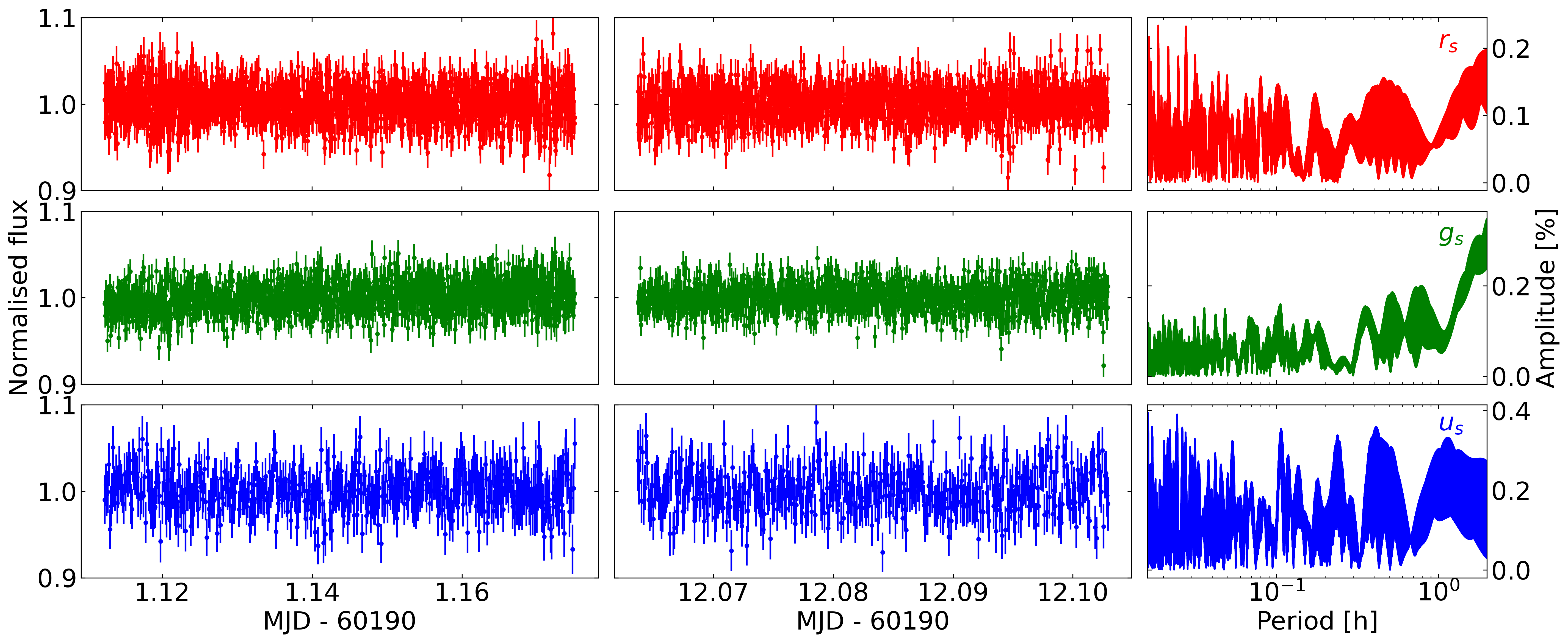}
    \caption{Light curves for 2023 September 03 (left panels) and 2023 September 14 (middle panels), and periodograms (right panels) for the $r_s$, $g_s$, and $u_s$ ULTRACAM data obtained for WD~J204259.71+152108.06. We find no evidence for periodic or aperidic variability.}
    \label{fig:j2042ucam}
\end{figure*}


\bsp	
\label{lastpage}
\end{document}